\documentclass[fleqn,usenatbib]{mnras}

\usepackage[T1]{fontenc}

\DeclareRobustCommand{\VAN}[3]{#2}
\let\VANthebibliography\thebibliography
\def\thebibliography{\DeclareRobustCommand{\VAN}[3]{##3}\VANthebibliography}

\usepackage{graphicx}	% Including figure files
\usepackage{mathrsfs}
\usepackage{amsfonts}
\usepackage{amstext}
\usepackage{amsmath}
\usepackage{amssymb}
\usepackage{hyperref}
\usepackage{tabularx}
\usepackage{upgreek}

\newcommand{\kms}{\ensuremath{\mathrm{km}\,\mathrm{s}^{-1}}}

\newcommand{\Lsun}{\ensuremath{\mathrm{L}_{\odot}}}
\newcommand{\Msun}{\ensuremath{\mathrm{M}_{\odot}}}

\newcommand{\XCO}{\ensuremath{X_\mathrm{CO}}}
\newcommand{\hi } {{\rm H}\,{\small\rm I}}

\usepackage{newtxtext,newtxmath}

\title[Feeding the SMBH in Fairall 49]{WISDOM Project - XIII. Feeding molecular gas to the supermassive black hole in the starburst AGN-host galaxy Fairall 49}

    \author[F. Lelli et al.]{Federico Lelli$^{1,2,}$\thanks{E-mail: federico.lelli@inaf.it}, Timothy A. Davis$^2$, Martin Bureau$^{3,4}$, Michele Cappellari$^3$, Lijie Liu$^3$, Ilaria Ruffa$^{2,5}$,\newauthor Mark D. Smith$^{3}$, Thomas G. Williams$^6$\\
$^1$INAF, Arcetri Astrophysical Observatory, Largo Enrico Fermi 5, I-50125, Florence, Italy\\
$^2$School of Physics and Astronomy, Cardiff University, Queens Buildings, The Parade, Cardiff, CF24 3AA, UK\\
$^3$Sub-department of Astrophysics, Department of Physics, University of Oxford, Keble Road, Oxford, OX1 3RH, UK\\
$^4$Yonsei Frontier Lab and Department of Astronomy, Yonsei University, 50 Yonsei-ro, Seodaemun-gu, Seoul 03722, Republic of Korea\\
$^5$INAF - Istituto di Radioastronomia, via P. Gobetti 101, 40129 Bologna, Italy\\
$^6$Max Planck Institut f{\"u}r Astronomie, K{\"o}nigstuhl 17, 69117 Heidelberg, Germany\\}

\date{Accepted 2022 August 31. Received 2022 August 30; in original form 2022 March 14}

\pubyear{2022}

\begin{document}
\label{firstpage}
\pagerange{\pageref{firstpage}--\pageref{lastpage}}
\maketitle

\begin{abstract}
The mm-Wave Interferometric Survey of Dark Object Masses (WISDOM) is probing supermassive black holes (SMBHs) in galaxies across the Hubble sequence via molecular gas dynamics. We present the first WISDOM study of a luminous infrared galaxy with an active galactic nuclei (AGN): Fairall\,49. We use new ALMA observations of the CO($2-1$) line with a spatial resolution of $\sim$80 pc together with ancillary HST imaging. We reach the following results: (1) The CO kinematics are well described by a regularly rotating gas disk with a radial inflow motion, suggesting weak feedback on the cold gas from both AGN and starburst activity; (2) The dynamically inferred SMBH mass is $1.6\pm0.4\mathrm{(rnd)}\pm0.8 \mathrm{(sys)}\times 10^{8}$\,\Msun\, assuming that we have accurately subtracted the AGN and starburst light contributions, which have a luminosity of $\sim$10$^9$ L$_\odot$; (3) The SMBH mass agrees with the SMBH$-$stellar mass relation but is $\sim$50 times higher than previous estimates from X-ray variability; (4) The dynamically inferred molecular gas mass is 30 times smaller than that inferred from adopting the Galactic CO-to-H$_2$ conversion factor ($\XCO$) for thermalised gas, suggesting low values of \XCO; (5) the molecular gas inflow rate increases steadily with radius and may be as high as $\sim$5 $M_\odot$\,yr$^{-1}$. This work highlights the potential of using high-resolution CO data to estimate, in addition to SMBH masses, the X$_{\rm CO}$ factor and gas inflow rates in nearby galaxies.
\end{abstract}

\begin{keywords}
black hole physics -- galaxies: active -- galaxies: ISM -- galaxies: kinematics and dynamics -- galaxies: Seyfert -- galaxies: starburst
\end{keywords}

%%%%%%%%%%%%%%%%%%%%%%%%%%%%%%%%%%%%%%%%%%%%%%%%%%

%%%%%%%%%%%%%%%%% BODY OF PAPER %%%%%%%%%%%%%%%%%%

\section{Introduction}

Supermassive black holes (SMBHs) with masses of $10^{6}-10^{9}$ M$_{\odot}$ are thought to reside at the center of every massive galaxy with stellar mass above a few times 10$^{9}$ M$_\odot$. The SMBH mass ($M_{\bullet}$) correlates with several galaxy properties, such as the total stellar mass ($M_\star$), the stellar mass of the bulge component ($M_{\rm bul}$), and the central stellar velocity dispersion ($\sigma_\star$) that is generally set by the dominant baryons in the core of massive galaxies \citep[e.g.,][]{Magorrian1998, Ferrarese2000, Marconi2003, Haring2004}. This evidence led to a general picture of galaxy-SMBH co-evolution, which remains debated and poorly understood \citep[see][for a review]{KormendyHo2013}. For example, despite SMBHs typically constituting only 0.1$\%$ of the host mass, they can release large amounts of energy during gas accretion phases, shining as active galactic nuclei (AGN) and impacting the host galaxy via positive or negative feedback \citep[e.g.,][]{Fabian2012, Morganti2018}.

A key observational issue concerns the measurement of the SMBH mass.  The most direct and reliable estimates of $M_\bullet$ are obtained using spatially resolved dynamics, probing the SMBH sphere of influence (SOI). The latter can be approximated as $r_{\rm SOI} = GM_{\bullet}/\sigma_{\star}^2$, where $\sigma_{\star}$ is a proxy for the inner stellar gravitational potential and $G$ is Newton's constant. For nearby galaxies with $M_\star\simeq10^{10-11}$ $M_\odot$, $r_{\rm SOI}$ ranges from a few tens to a few hundreds of pc, so it can be resolved using observations of stellar kinematics, ionized gas kinematics, and megamaser kinematics. A proper determination of the SOI of the SMBH, however, requires a full calculation of the stellar gravitational potential (rather than simply using $\sigma_\star$), so the kinematic data need to be supplemented with high-resolution surface photometry that traces the distribution of the stellar mass.

\begin{figure*}
	\includegraphics[width=0.475\textwidth]{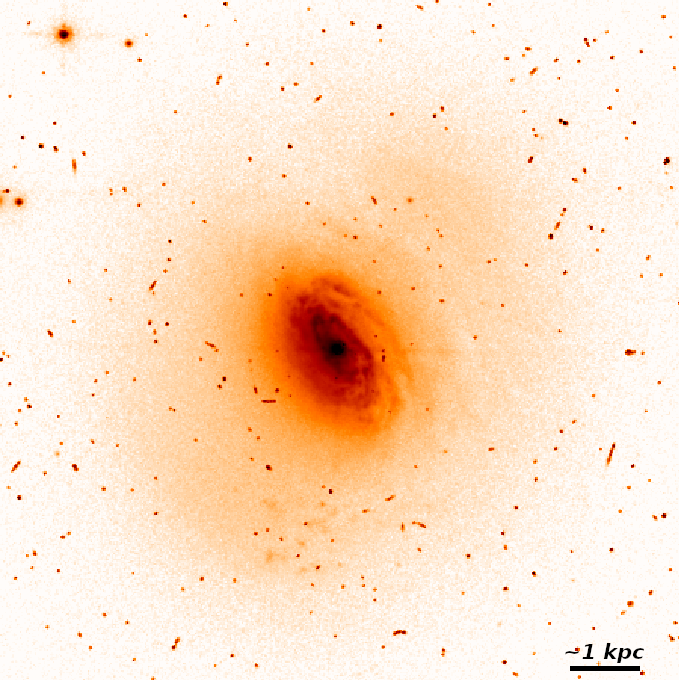}
	\includegraphics[width=0.475\textwidth]{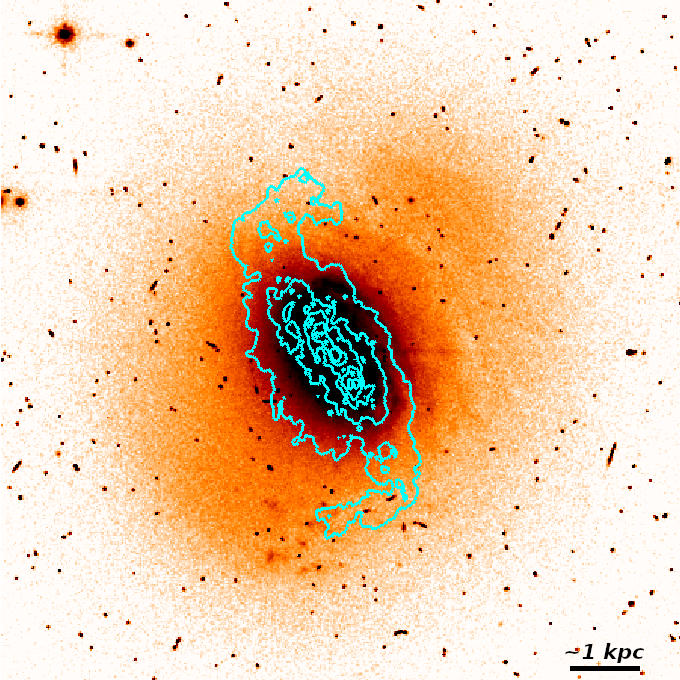}
    \caption{HST/WFPC2 image of Fairall\,49 in the F606W filter. The two panels show the same HST image but use a different color stretching to highlight different features in the galaxy. The left panel shows the inner spiral disk, while the right panel highlights the outer asymmetric stellar distribution superimposed with the CO(2-1) emission (cyan contours). Contours increase from 0.14 ($\sim$3$\sigma$) to 8 mJy beam$^{-1}$ km s$^{-1}$. The bar to the bottom-right corner equals $\sim$2.5$''$, corresponding to $\sim$1 kpc for the assumed distance of 86.7 Mpc. North is up; East is left.}
    \label{fig:HST}
\end{figure*}

Each one of the kinematic tracers mentioned above have unavoidable pros and cons \citep[see][for details]{KormendyHo2013}. In general, measurements based on stellar and ionized gas kinematics need to account for both rotation and pressure support, which is non-trivial due to possible anisotropies in the velocity dispersion tensor. Megamasers, instead, involve dynamically cold gas disks in which the pressure support is negligible, but megamasers are rare so this technique cannot be applied to large, representative galaxy samples. Over the past decade, another method has become available thanks to the substantial increase in angular resolution and sensitivity of sub-millimeter interferometry: probing the rotation of dynamically cold gas disks using CO emission lines \citep{Davis2013}. 

The mm-Wave Interferometric Survey of Dark Object Masses (WISDOM) project is measuring SMBH masses using high-resolution CO observations from the Atacama Large Millimeter/submillimeter Array (ALMA) and the Combined Array for Research in Millimeter-wave Astronomy (CARMA). The WISDOM sample probes a morphologically diverse sample of galaxy hosts, including both early-type galaxies (ETGs: ellipticals and lenticulars) and late-type galaxies (LTGs: spirals and irregulars). To date, the WISDOM project has provided SMBH masses in six typical ETGs \citep{WISDOM-1, WISDOM-2, WISDOM-3, WISDOM-4, WISDOM-5, WISDOM-7} and even a dwarf ETG \citep{Davis2020}. Similar molecular-gas measurements have been presented by different teams for another four ETGs \citep{Barth2016b, Boizelle2019, Nagai2019, Ruffa2019}, seven LTGs with AGNs \citep{Combes2019}, and a barred spiral galaxy \citep{Nguyen2020}. The WISDOM project has also investigated the scaling relation between $M_\bullet$ and CO linewidths \citep{WISDOM-6}, the molecular gas morphologies in galaxy centers \citep{WISDOM-10}, the feedback processes in the brightest cluster galaxy NGC\,708 \citep{WISDOM-8}, and the properties of giant molecular clouds in the lenticular galaxy NGC\,4429 \citep{WISDOM-9}.

This paper presents the first WISDOM study of a peculiar object: the luminous infrared galaxy (LIRG) Fairall\,49, which hosts a central AGN and appears to be a late-stage merger (see Figure\,\ref{fig:HST}). We start by reviewing the properties of Fairall\,49 (Sect.\,\ref{sec:target}). Then, we present new ALMA observations and ancillary data (Sect.\,\ref{sec:data}) that are used to build dynamical models (Sect.\,\ref{sec:dynamics}). Next, we discuss the implications of our results for the SMBH-galaxy co-evolution, the CO-to-H$_2$ conversion factor, and the baryon cycle in AGNs (Sect.\,\ref{sec:discussion}). Finally, we provide a concise summary (Sect.\,\ref{sec:summary}).

\section{The target: a LIRG with a Seyfert nucleus}\label{sec:target}

\citet{Fairall1977} used the European Southern Observatory (ESO) Quick Blue Survey to discover 150 galaxies with regions of exceptionally high surface brightness (HSB), including Fairall\,49. This object was later associated with the IR-detected galaxy IRAS 18325–5926 \citep{Carter1984} and with an X-ray selected AGN \citep{Piccinotti1982, Ward1988}. The optical classification of the AGN has been debated: \citet{Osterbrock1985} and \citet{deGrijp1985} classified the galaxy as a Seyfert-2 (Sy2, i.e., without a broad line region), while \citet{Carter1984} and \citet{Maiolino1995} proposed a Sy1.8 type (i.e., with weak broad lines in H$\alpha$ and H$\beta$ emission). Finally, \citet{VeronCetty2006} gave a Sy1h classification: a Sy2 spectra that shows a Sy1 behaviour in polarized light. The X-ray emission from the AGN shows flux variability \citep{Iwasawa1995} and a broad Iron Fe k emission line \citep{Iwasawa1996, Iwasawa2016}.

We estimate the distance ($D$) of Fairall\,49 using the Cosmiflows-3 distance-velocity calculator \citep{Kourkchi2020}, which computes distances from line-of-sight velocities considering the peculiar velocity field of the nearby Universe \citep{Graziani2019}. This flow model is consistent with a Hubble constant ($H_0$) of 75 $\kms$ Mpc$^{-1}$ \citep{Kourkchi2020, Schombert2020}. Considering that Fairall\,49 has a line-of-sight velocity of 6002 \kms \citep{Huchra2012}, we obtain $D = 86.7$ Mpc, so 1$''$ corresponds to $\sim$0.42 kpc. For comparison, the NASA/IPAC Extragalactic Database (NED)\footnote{\url{https://ned.ipac.caltech.edu/}} gives a distance of $88.1\pm6.2$ Mpc using the velocity frame set by the Cosmic Microwave Background. For large line-of-sight velocities, the effect of peculiar velocities on $D$ is not large, so the distance uncertainty is plausibly of the order of 10$\%$. This translates to a systematic error of $\sim$10$\%$ on dynamics-based masses and $\sim$20$\%$ on luminosity-based ones. Table\,\ref{tab:basic} provides the basic parameters of the galaxy.

\begin{table}
\caption{Basic Properties of Fairall\,49}
\label{tab:basic}
\centering
\begin{tabular}{l c c}
\hline
Property                  & Value & Reference \\
\hline
Right Ascension (deg.)    & 279.242875 & NED \\
Declination (deg.)        & $-$59.402389 & NED \\
Redshift                  & 0.02002   & \citet{Huchra2012}\\
$D$ (Mpc)                 & 86.7      & This work (Sect.\,\ref{sec:target})\\
$M_{F606W}$ (mag)         & $-20.31$ & This work (Sect.\,\ref{sec:photo})\\
$R_{\rm eff}$ (pc)        & 882       & This work (Sect.\,\ref{sec:photo})\\ 
$L_{F606W}$ (L$_\odot$)   & $1.2\times10^{10}$ & This work (Sect.\,\ref{sec:photo})\\
$L_{\rm FIR}$ (L$_\odot$) & $1.1 \times 10^{11}$ & \citet{MunozMarin2007}\\
$L_{\rm X}$ (erg s$^{-1}$)          & $1.5\times10^{43}$      & \citet{Iwasawa2016}\\
SFR ($M_\odot$ yr$^{-1}$)    & 16.4 & This work (Sect.\,\ref{sec:target})\\
\hline
\end{tabular}
\end{table}
Fairall\,49 has a total far-IR luminosity of $1.1 \times 10^{11}$ \Lsun\, \citep[from IRAC fluxes; see][]{MunozMarin2007}, so it fits within the class of luminous infrared galaxies \citep[LIRGs,][]{SandersMirabel1996}. Assuming that the far-IR luminosity traces dust heated by young stars, we obtain a star-formation rate (SFR) of 16.4 M$_\odot$ yr$^{-1}$ using the calibration from \citet{Murphy2011}. This is just a rough order-of-magnitude estimate of the SFR because the far-IR luminosity can be affected by the AGN emission \citep{SandersMirabel1996}. The stellar mass of Fairall\,49 is about  $2\times10^{10}$ M$_\odot$ (see Sect.\,\ref{sec:mass} and Table\,\ref{tab:MCMC}), so this object is above the ``main sequence'' of star-forming galaxies \citep[e.g.][]{Renzini2015, McGaugh2017}. Most likely, the AGN co-exists with a nuclear starburst.

The morphological classification of Fairall\,49 is unclear. Hyperleda\footnote{\url{http://leda.univ-lyon1.fr/}} \citep{Makarov2014} classifies the galaxy as E$-$S0 ($T=-3\pm2$): this classification is most natural if one visually inspects the available ground-based imaging. However, \citet{Malkan1998} and \citet{MunozMarin2007} report a Sa classification based on \emph{Hubble Space Telescope} (HST) imaging. Figure\,\ref{fig:HST} (left) shows the HST image in the F606W filter: the galaxy has tightly wound spiral arms within $\sim$5$''$ ($\sim$2 kpc) and a compact light concentration within $\sim$0.5$''$ ($\sim$0.2 kpc). The inner light concentration is likely dominated by AGN and starburst emission, thus it cannot be naively interpreted as a bulge. The light concentration is also slightly offset ($\sim$100 pc) from the outer isophotes, possibly due to dust obscuration. Figure\,\ref{fig:HST} (right) highlights the outer light distribution, which displays a complex and asymmetric morphology with shells and knots reminiscent of past interaction events. Moreover, the CO(2-1) emission (from our new ALMA data, Sect.\,\ref{sec:ALMA}) shows two outer spiral arms and/or tidal tails, which may also point to a past interaction.  Thus, it is tempting to reclassify Fairall\,49 as a Sc galaxy with an outer, merger-driven stellar component. Alternatively, the galaxy may be a dust-obscured elliptical with a central spiral disk, possibly due to a recent gas accretion event. The half-light radius from HST photometry (Sect.\,\ref{sec:photo}) is 2.1$''$ ($\sim$0.9 kpc) and the visible radial extent is around $14''$ ($\sim$6 kpc). These properties are comparable to local early-type galaxies \citep[e.g., ][]{Kormendy2012} as well as to massive starburst and post-starburst galaxies at high $z$ \citep[e.g.][]{Almaini2017}. Observations with integral-field spectrographs are needed to study the stellar kinematics of Fairall\,49 and shed new light on the stellar component.

\section{Data analysis}\label{sec:data}

\subsection{ALMA observations}\label{sec:ALMA}

Fairall\,49 was observed as part of project 2017.1.00904.S (PI: M. Smith) with two different configurations of the ALMA 12-m array: (i) the relatively compact configuration C43-5 on 17 September 2018 with 44 antennas, providing minimum and maximum baselines of 14 and 1397 m, respectively; (ii) the more extended configuration C43-8 on 17 July 2019 with 43 antennas, providing minimum and maximum baselines of 91 m and 8547 m, respectively. The flux and bandpass calibrator was J1924-2914 for C43-5 and J2056-4714 for C43-8, while the phase calibrator was J1829-5813 for both configurations. The time on source was about 5 minutes for both execution blocks. Two additional execution blocks were obtained in configuration C43-8 on 23 and 26 November 2017, but they were flagged as ``semi-pass'' by the ALMA quality-assurance team and are not analysed here.

We used ALMA band 6 with a mixed spectral setup. A high-resolution spectral window with a bandwidth of 1.875 GHz was covered with 3840 channels and centered at a rest-frequency of 230.538 GHz to target the CO$(J=2\rightarrow1)$ emission line. Three low-resolution spectral windows with bandwidths of 1.875 GHz were covered with 128 channels and centered at 232.5, 246.0, and 248.0 GHz to target the dust-emitting continuum.

The data reduction was performed with version 5.6.1-8 of the Common Astronomy Software Applications (\textsc{Casa}) package \citep{CASA}. The $uv$ data were flagged and calibrated using the standard \textsc{Casa} pipeline. After preliminary imaging of the combined dataset, additional bad antennas and baselines were visually identified and flagged using the CASA task {\tt flagdata} in the {\tt manual} mode. The percentage of data flagged using this strategy was only a few percent, so it does not significantly affect the quality of our final imaging products. Both the line and continuum data were imaged using the \texttt{tclean} task with a H\"ogbom deconvolver and Briggs weighting with a robust parameter of 0.5. The images were interactively cleaned down to $\sim$3$\sigma$. The continuum image was obtained combining all four spectral windows, excluding channels with line emission. The synthesized beam of the continuum map is $0.171'' \times 0.132''$ ($72 \times 55$ pc) with a position angle (PA) of $9.52^{\circ}$, and the root-mean-square (rms) noise is 0.038 mJy beam$^{-1}$.

The CO datacube was obtained after subtracting the continuum in the $uv$ plane using the \textsc{Casa} task \texttt{uvcontsub}, fitting a first-order polynomial to the line-free channels. The spectral channels were binned to 15 \kms to increase the signal-to-noise ratio. The channel width was chosen as the best compromise between spectral resolution and sensitivity. Two cleaned cubes were created testing two different cleaning approaches: (1) using the task \texttt{tclean} in \textsc{Casa} in interactive mode, identifying by eye spatial regions of line emission in each channel, (2) using the task \texttt{clean} in the Groningen Imaging Processing System (\textsc{Gipsy}) package\footnote{ \url{https://www.astro.rug.nl/~gipsy/}} \citep{Gipsy}, adopting a Boolean mask that was constructed by smoothing the dirty cube to a resolution of 0.4$''$ and clipping at 3$\sigma_{\rm s}$ (where $\sigma_{\rm s}$ is the rms noise of the smoothed cube). The results of our dynamical modeling (Sect.\,\ref{sec:dynamics}) do not strongly depend on the adopted cleaning procedure. In the following, we use the cleaned cube from \textsc{Gipsy}. The median synthesized beam of the CO cube is $0.216''\times 0.160''$ ($91 \times 67$ pc) with a PA of $3.95^{\circ}$. Moment maps were constructed using the \textsc{$^{\rm 3D}$Barolo} package\footnote{\url{https://bbarolo.readthedocs.io/en/latest/}} \citep{Barolo}. In particular, we consider the signal inside a Bolean mask constructed with the task \texttt{smooth \& search} (with default values). We also built a map of the mean rms noise at each spatial position and checked that it is Gaussian with a mean value of 0.7 mJy/beam per channel.

\begin{figure}
	\includegraphics[width=0.48\textwidth]{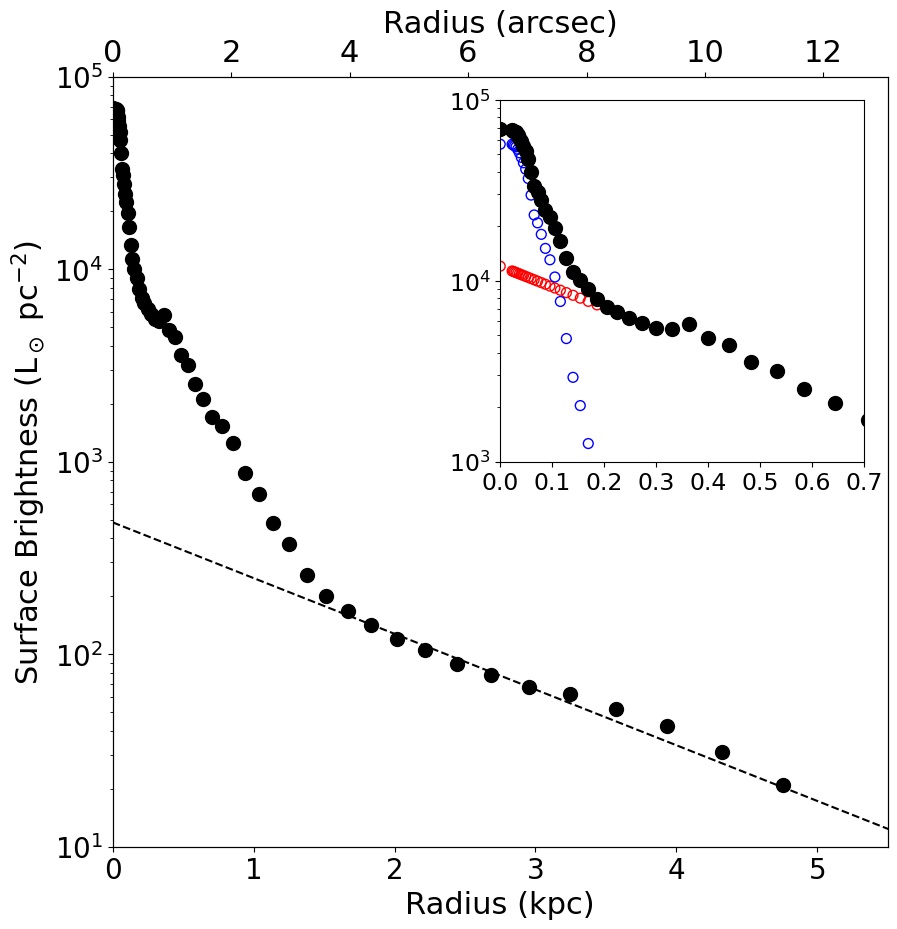}
    \caption{Surface brightness profile of Fairall\,49 from the HST/WFPC2 image in the F606W filter. The dashed line shows an exponential fit to the outer parts. The inset zooms at small radii ($R<0.7$ kpc): the blue circles corresponds to the inner HSB component that is likely dominated by the AGN and/or starburst emission, the red circles shows the inward exponential extrapolation of the stellar component. See text for details.}
    \label{fig:sbp}
\end{figure}
\subsection{HST surface photometry}\label{sec:photo}

Several images of Fairall\,49 are available in the Hubble Legacy Archive\footnote{\url{https://hla.stsci.edu/}}. Since we are interested in modeling the stellar mass distribution and calculating the stellar gravitational field, we analysed the reddest optical image available, taken with the F606W filter of the Wide Field and Planetary Camera 2 (WFPC2; Project ID 5479). A near-IR image in the F160W filter of the Near Infrared Camera and Multi-Object Spectrometer (NICMOS) exists but is heavily saturated, so is not useful for our purposes. We also analysed a ground-based $J$-band image from the VISTA Hemisphere Survey \citep{VHS} but its angular resolution is too low to trace the inner light distribution and the gain in surface brightness in the outer parts is very modest with respect to the HST photometry, thus we will not discuss this image further.

We performed surface photometry on the Planetary Camera (PC) frame (Fig.\,\ref{fig:HST}) using the \textsc{Archangel} software\footnote{\url{ http://abyss.uoregon.edu/~js/archangel/}} \citep{Archangel}. The sky background was determined selecting boxes with no galaxy emission. Foreground stars and background galaxies were interactively masked. Next, ellipses were automatically fitted to the galaxy isophotes and the azimuthally-averaged surface brightness profile was determined. We also built the luminosity curve-of-growth, replacing empty pixels with the mean value over the isophote, and determined the total magnitude with an asymptotic fit. The results are given in Table\,\ref{tab:basic}. The weak horizontal stripe in the middle of Fig.\,\ref{fig:HST} shows that the innermost pixels are saturated, but their contribution to the total luminosity is negligible (a few percent).

\begin{figure*}
	\includegraphics[width=\textwidth]{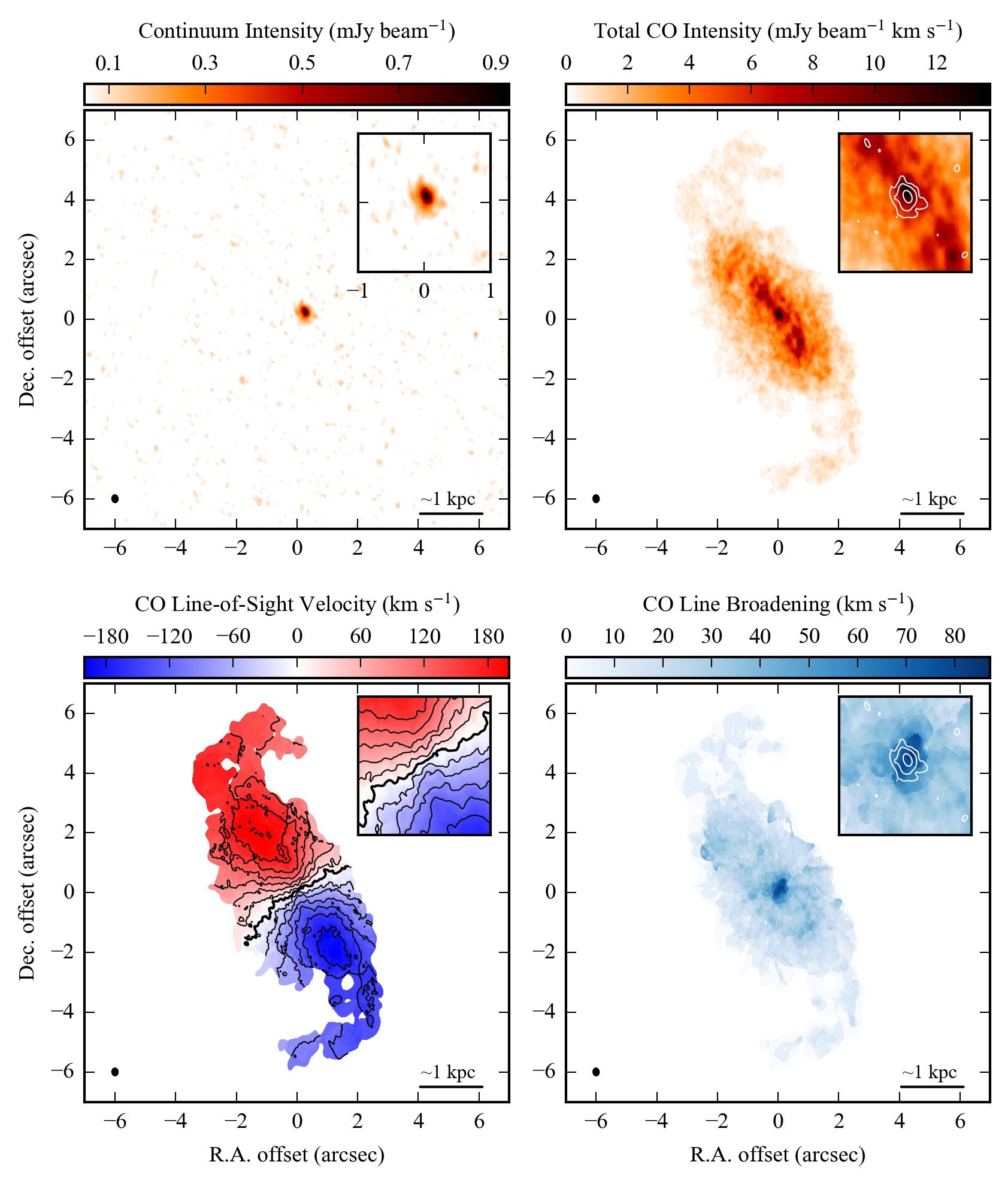}
    \caption{Overview of ALMA observations: 1.3 mm continuum map (top-left panel), CO moment-zero map (top-right panel), CO moment-one (bottom-left panel), and CO moment-two map (bottom-right panel). In all panels, the ellipse in the bottom-left corner shows the synthesized beam, while the bar in the bottom-right corner corresponds to $\sim$1 kpc. The insets zoom into the central $2''\times 2''$ area; the moment-zero and moment-two maps are overaild with the 1.3 mm continuum emission (white contours at 3$\sigma$, 6$\sigma$, and 12$\sigma$). In the moment-one map, iso-velocity contours range from $\pm240$ \kms\ in steps of 30 \kms\ (twice the channel width); the bold contour corresponds to the systemic velocity (set to zero).}
    \label{fig:maps}
\end{figure*}

The surface brightness profile (Figure\,\ref{fig:sbp}) shows a complex shape with three main regimes: (1) a very HSB region within 0.2 kpc ($R \lesssim 0.5''$) that is likely dominated by AGN and starburst emission; (2) an exponential component between 0.2 and 1.5 kpc ($0.5'' \lesssim R \lesssim 3.5''$) with a kink at 0.4 kpc that corresponds to tightly wounded spiral arms forming a ring-like structure (Fig.\,\ref{fig:HST}, left panel), and (3) a nearly exponential extension beyond 1.5 kpc  ($R\gtrsim3.5''$) obtained by averaging over the asymmetric low-surface-brightness stellar emission (Fig.\,\ref{fig:HST}, right panel). For the purpose of comparing with other galaxies, we fitted the data at $R>1.5$ kpc with an exponential function, giving a central surface brightness $\mu_0 = 19.8$ mag/arcsec$^2$ and scale length $R_{\rm exp} = 1.5$ kpc ($\sim3.6''$). These values are comparable with the most compact galaxy disks at $z\simeq0$ \citep{vanDerKruit2011} but we stress that the morphology of the outer light component does not resemble a regular stellar disk.

\subsection{Stellar gravitational potential}\label{sec:stellarpot}

In previous WISDOM papers, the stellar gravitational contribution to the circular velocity was computed using the \textsc{MgeFit} software\footnote{\url{https://www-astro.physics.ox.ac.uk/~mxc/software/}} \citep{Cappellari2002}, which fits 2-dimensional (2D) images with the Multi-Gaussian Expansion (MGE) method \citep{Emsellem1994}. In the case of Fairall\,49, the 2D \textsc{MgeFit} gives a stellar velocity curve that completely differs in shape from the observed rotation curve derived in Sect.\,\ref{sec:Barolo}, unless we remove the two innermost Gaussian components from the calculation. This suggests that the inner HSB component is dominated by AGN and starburst emission, rather than tracing stellar mass. Importantly, the inner HSB component is resolved by the HST point-spread function (PSF), so we cannot merely perform a PSF subtraction of a nuclear point source in 2D to solve the issue. Thus, we prefer to compute the stellar gravitational contribution using the azimuthally-averaged surface brightness profile (Fig.\,\ref{fig:sbp}), performing a classic 1-dimensional (1D) subtraction of the HSB component to have more direct control on the inner light profile.

The HSB component is decomposed in a non-parametric fashion, following a similar strategy as \citet{Lelli2016} for bulge-disk decompositions. In short, we identify a radius within which the AGN and/or starburst emission dominates the optical light, which we call $R_{\rm HSB}$. Next we extrapolate an exponential disk for $R < R_{\rm HSB}$ and subtract it from the total luminosity profile. The resulting HSB component is extrapolated for $R>R_{\rm HSB}$ and the procedure is iterated in such a way that the sum of the two components conserve the total surface brightness at each radius. The result is shown by the inset in Fig.\,\ref{fig:sbp} for our fiducial light subtraction. Different light decompositions are discussed in Appendix\,\ref{sec:sys}. For our baseline model, the luminosity of the HSB component is about $10^9$ L$_\odot$ within $\sim$200 pc ($\sim$0.5$''$), corresponding to 8$\%$ of the total luminosity of Fairall\,49. If we use the AGN templates of \citet{Stalevski2012, Stalevski2016} scaled to the bolometric luminosity of Fairall\,49 ($\sim$10$^{11}$ from X-ray observations) and assume that the AGN torus is co-aligned with the disk, we find that the F606W luminosity of the AGN is in the range 10$^{8}-10^{9}$\,L$_\odot$, consistent with the luminosity of the HSB component. 

We explored two different approaches to compute the stellar gravitational contribution from the 1D luminosity profile (with or without the HSB component). The first approach (1D MGE) fits the surface brightness profiles with \textsc{MgeFit} \citep{Cappellari2002} approximating the profile with multiple Gaussians, then the gravitational potential of each Gaussian is computed using the equation from \citet{Chandra1969} for the gravitational potential of a density distribution on concentric homeoids. The second approach (``thick disk'') uses the surface brightness profile as input to solve the equation from \citet{Casertano1983} for the gravitational potential of an axisymmetric disk with constant thickness\footnote{We solve Casertano's equation using the \textsc{Gipsy}'s task \texttt{Rotmod}.}. Although both approaches reproduce the same 1D surface brightness profile, they assume different vertical density distributions. In the 1D MGE approach, we adopt a constant axial ratios $q=0.01$ for all the MGE Gaussian components, so the density is stratified on oblate spheroids. The adopted $q=0.01$ is so small that the MGE model effectively approximates an infinitely thin disk, giving realistic, self-similar elliptical isophotes in 2D projection. In the thick-disk approach, we assume an exponential vertical density distribution with fixed scale height $h_{\rm z} = 253$\,pc at each radius. This scale height is motivated by the empirical relation $h_{\rm z} = 0.196 h_{\rm R}^{0.633}$ from edge-on disk galaxies \citep{Bershady2010}, where $h_{\rm R}$ is the observed scale length in the outer regions ($\sim$1.5 kpc for Fairall\,49).

The results from the two approaches are described in detail in Appendix\,\ref{sec:sys}. Importantly, both approaches point to the need of subtracting a specific amount of inner light ($R<200$ pc), but for different reasons. For models with a weak or null HSB light subtraction, the 1D MGE approach cannot reproduce the observed rotation curve (similarly to the 2D MGE results), whereas the thick-disk approach cannot reproduce a sensible projected 2D image. The latter issue occurs because, when the steepness of the radial surface brightness profile is larger than the assumed vertical steepness, the projected isophotes of the model image become oriented perpendicular to the projected major axis of the stellar disk. Thus, these models have effectively prolate nuclei, which are qualitatively inconsistent with the observed shapes of most galaxy nuclei. Moreover, these prolate nuclei generate a reduced radial gravitational force in the disk plane, so they provide smaller circular velocities than either a spherical-like or a disk-like density distribution. Ironically, this effect allows the thick-disk models to fit the observed rotation curve irrespective of how we treat the inner HSB component, thus they can be used to estimate conservative systematic uncertainties due to AGN and starburst contributions as well as 3D geometry (see Appendix \ref{sec:sys}).

Overall, both approaches indicate that models with a substantial HSB component are unable to fit the data, so they must be discarded. For our adopted subtraction of the inner HSB component, both the thick-disk and the 1D MGE models generate similar circular velocities and can reproduce both kinematic and photometric data. This gives us confidence that our conclusions are not strongly affected by the adopted approach, nor by the level of light subtraction. In the following, we report the results from the thick-disk model but this does not affect our general conclusions.

\begin{figure*}
	\includegraphics[width=1.0\textwidth]{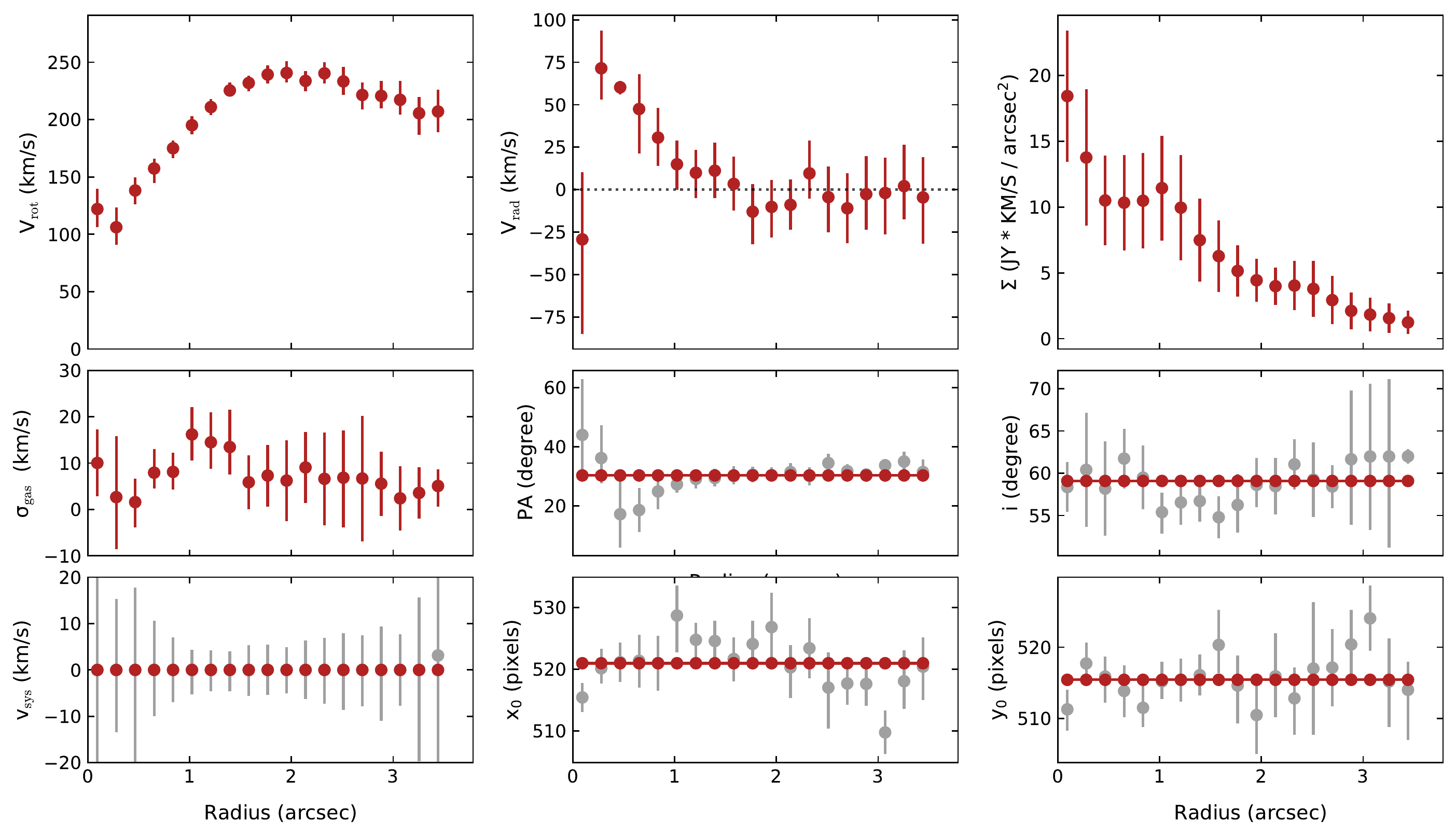}
    \caption{Best-fit parameters from \textsc{$^{\rm 3D}$Barolo}. From top to bottom, from left to right: rotation velocity, radial velocity, surface brightness profile, velocity dispersion profile, position angle, inclination angle, systemic velocity, coordinates of the kinematic center. Grey points show initial estimates from a first fit in which all parameters are free, while red points show final estimates from a second fit in which $i$, PA, $V_{\rm sys}$, $x_0$ and $y_0$ are fixed to the mean value across the rings.}
    \label{fig:BB}
\end{figure*}

\section{Results}\label{sec:dynamics}

\subsection{Gas distribution \& kinematics}

Figure \ref{fig:maps} provides an overview of the ALMA observations of Fairall\,49. The continuum emission at 1.3 mm is very compact and barely resolved at $\sim$0.15$''$ resolution. The CO$(J=2\rightarrow1)$ emission, instead, extends for $\sim$12$''$ ($\sim$5 kpc) and shows symmetric tails in the outer parts, reminiscent of spiral arms. The winding of these outer tails is similar to that of the inner spiral arms visible in the HST image (Fig.\,\ref{fig:HST}, left panel). The 1.3 mm continuum emission roughly coincides with the peak of the CO emission.

The CO velocity map reveals a rotating disk with a diameter of $\sim$6$''$ ($\sim$3 kpc). The presence of a regular CO disk is not trivial given the LIRG and Seyfert nature of Fairall\,49: the combined feedback from AGN and starburst activity appears to not strongly affect the overall kinematics of the molecular gas. Interestingly, the inner velocity contours ($R\lesssim1.2''$) display a symmetric distortion: the direction of maximal velocity gradient is not perpendicular to the systemic velocity but skewed. This is a classic signature of radial motions in the disk plane \citep[e.g.][]{Lelli2012a, Lelli2012b}. The outer gaseous arms/tails are kinematically connected with the inner disk. The gas kinematics become more complex at the base of the tails.

The CO velocity dispersion map shows enhanced line broadening at the position of the 1.3 mm continuum emission. The observed velocity dispersion, however, should be interpreted with caution because the channel width is non-negligible (15 \kms) and the CO line profile is broadened by unresolved motions within the synthesized beam (beam-smearing effects). In the following, we study the gas kinematics using 3D techniques that take the effects of both spectral and spatial resolution into account.

Dynamical models are built using two different approaches: (i) we fit a kinematic tilted-ring model to the cube (Sect. \ref{sec:Barolo}) using \textsc{$^{\rm 3D}$Barolo} \citep{Barolo}, then we fit a mass model to the rotation curve (Sects. \ref{sec:mcmc} and \ref{sec:mass}); and (ii) we fit a mass model directly to the datacube (Sect. \ref{sec:KinMS}) using \textsc{KinMS} \citep{KinMS, Davis2020b} in a forward-modeling fashion. Both approaches have pros and cons. Approach (i) is computationally fast and allows modeling the gas dynamics in a non-parametric fashion, but it is not straightforward to investigate possible degeneracies between the kinematic parameters. Approach (ii) is more computationally expensive and requires defying a parametric model (including the gravitational potential) but explores the full parameter space with a Markov-Chain-Monte-Carlo (MCMC). As we will see, both approaches give consistent results within the uncertainties. We start with describing the tilted-ring fits with \textsc{$^{\rm 3D}$Barolo} because their results are later used to guide the parametric modeling with \textsc{KinMS}.

\begin{figure*}
	\includegraphics[width=1.0\textwidth]{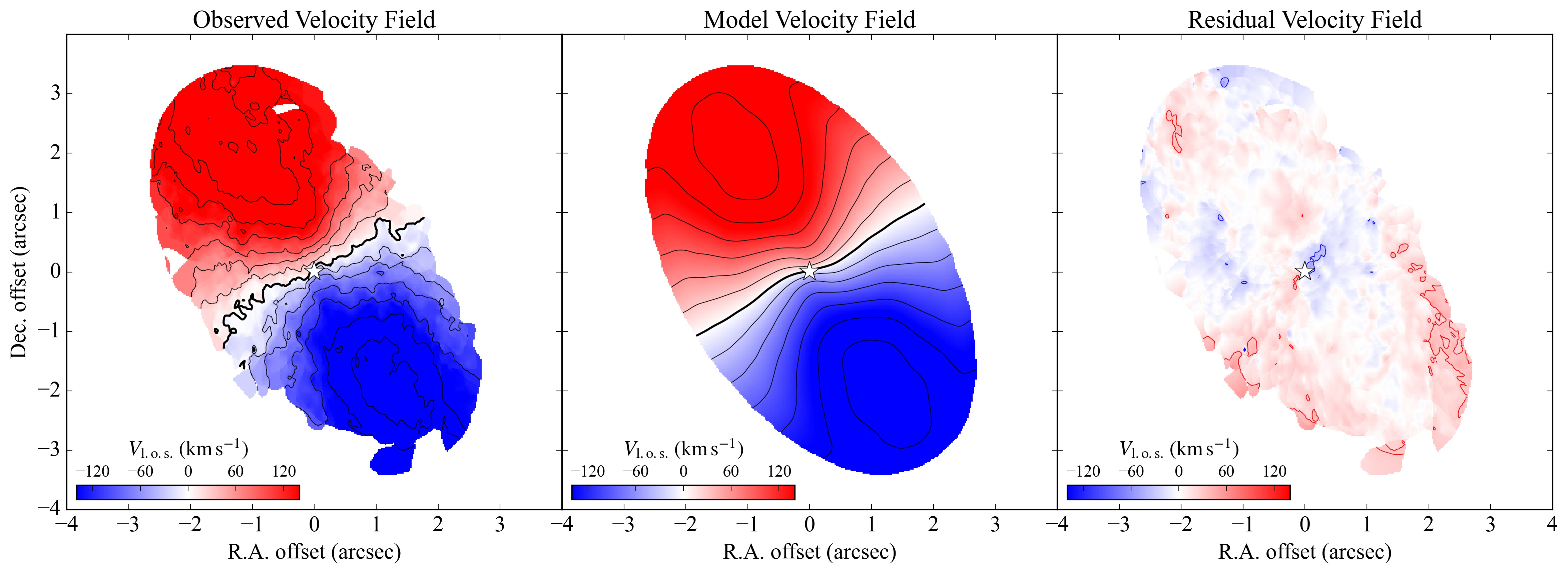}
    \caption{Velocity fields from the observed cube (left panel), model cube (middle panel) and residual cube (right panel). The velocity fields are cropped beyond $R\simeq3.5''$ because we do not model the kinematics at larger radii. In the left and middle panels, the velocity contours range from $\pm$240 \kms in steps of 30 \kms (twice the channel width). In the right panel, the velocity contours are at $\pm$30 \kms. In all panels, the white star corresponds to the dynamical center.}
    \label{fig:Maps}
\end{figure*}

\begin{figure}
	\includegraphics[width=0.5\textwidth]{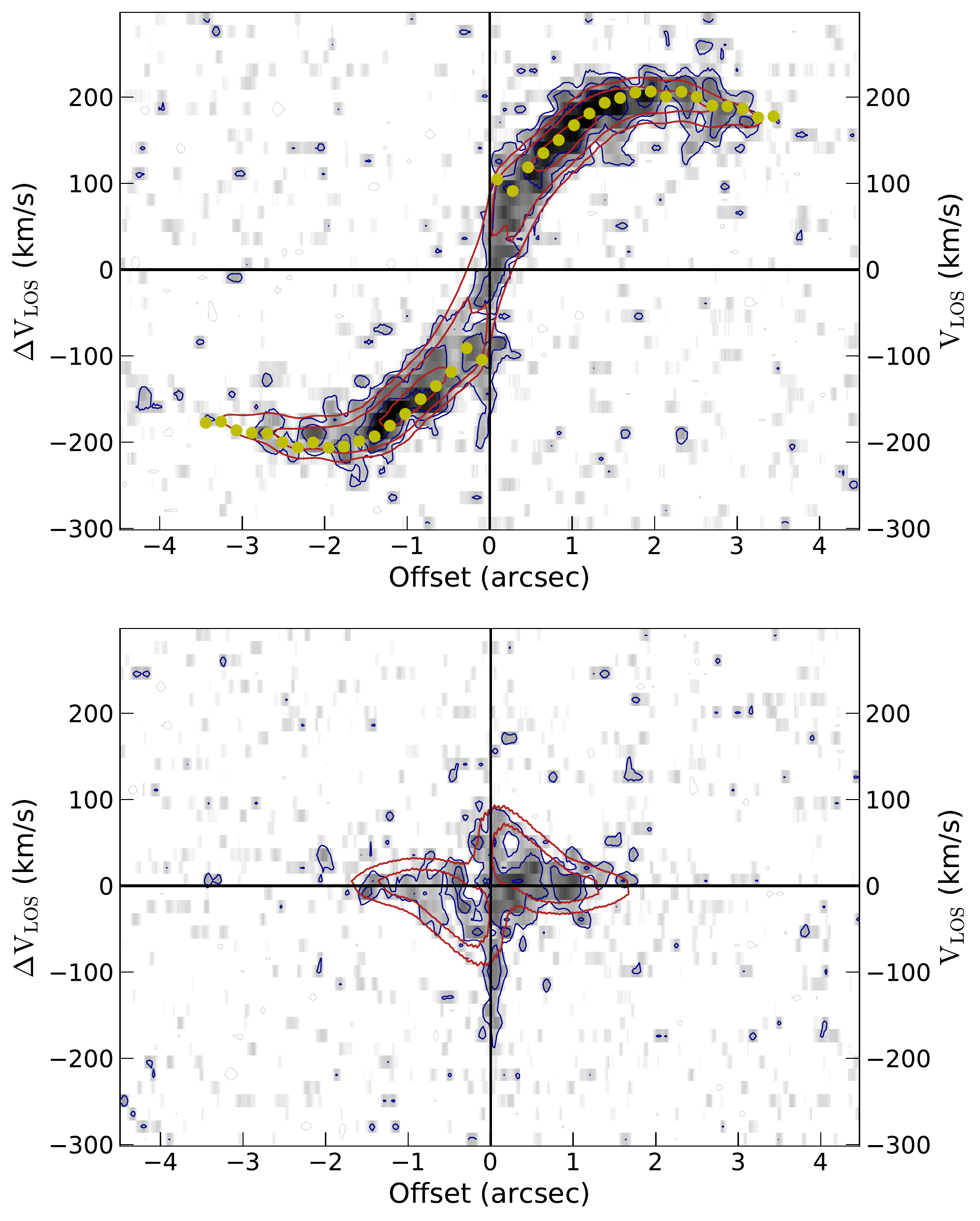}
    \caption{Position-velocity diagrams along the major (top) and minor (bottom) axes of the CO disk (see dashed lines in Figure\,\ref{fig:maps}). Positive and negative velocities correspond to the North-Eastern and South-Western sides of the disk, respectively (see Fig.\,\ref{fig:Maps})}. The data (greyscale) are overlaid with the best-fit model from \textsc{$^{\rm 3D}$Barolo} (red contours) and the projected rotation curve (yellow dots). Contours correspond to 2, 4, and 8$\sigma$ where $\sigma =  0.84$ mJy/beam is the rms noise.
    \label{fig:PV}
\end{figure}

\subsection{Tilted-ring modeling: rotation and non-circular motions}\label{sec:Barolo}

\textsc{$^{\rm 3D}$Barolo} divides the galaxy into a series of rings, where each ring is characterized by nine parameters. Four parameters concern the rings' geometry: center ($x_0, y_0$), systemic velocity ($V_{\rm sys}$), position angle (PA), and inclination angle ($i$). The other five parameters concern physical properties: surface density ($\Sigma_{\rm gas}$), vertical thickness ($z_0$), rotation velocity ($V_{\rm rot}$), radial velocity ($V_{\rm rad}$), and velocity dispersion ($\sigma_{\rm V}$). For simplicity, we adopt a fully axisymmetric disk, so the surface density of each ring is directly computed from the observed CO intensity map using azimuthal averages. We assume an exponential vertical distribution with a fixed scale height of 100 pc (0.25$''$): the precise value of $z_0$ has virtually no effects on our results. Thus, we are left with seven free parameters in each ring.

We adopt 19 rings with a width of 0.186$''$, corresponding to the geometric resolution element $\sqrt{a\times b}$ where $a$ and $b$ are the major and minor axes of the synthesized beam, thus the rings are fully independent. The radial extent of our disk model ($\sim3.5''$) excludes the outermost gaseous arms/tails, which we do not attempt to fit because they probably do not trace the galaxy circular velocity. Within each ring, the 3D pixels are weighted according to cos($\theta$) where $\theta$ is the azimuthal angle in the disk plane measured from the projected major axis; this is a standard approach in tilted-ring modeling because the kinematic information lies mainly along the disk major axis. We set the options ``twostage=true'' and ``polyn=0'': in a first run all seven parameters are left free, then the geometric parameters (center, $V_{\rm sys}$, PA, and inc) are fixed to the mean values across the rings and a second fit is performed with only three free parameters ($V_{\rm rot}$, $V_{\rm rad}$, $\sigma_{\rm V}$). Thus, we model the gas in this system as a flat disk because there are no strong indications for a warped disk, as we show below.

The best-fit parameters are shown in Figure\,\ref{fig:BB}. The value of $V_{\rm sys}=0$ \kms corresponds to a fiducial redshift $z=0.020021$ \citep{Huchra2012} or equivalently to $c \times z = 6002$ \kms, which we used in Sect.\,\ref{sec:target} to estimate the galaxy distance. The inner point of the rotation curve is higher than the second one, providing a first indication of the presence of a central SMBH, which will be confirmed in Sect.\,\ref{sec:KinMS} using \textsc{KinMS}. The intrinsic velocity dispersion varies from a few \kms up to $\sim$15 \kms but the errorbars are substantial. Radial motions are clearly detected at small radii ($R<1''$), while they are consistent with zero at larger radii. The overall gas kinematics, however, are dominated by rotation with $V_{\rm rot}> 1.5 \, V_{\rm rad}$ at all radii.

In general, radial motions may be degenerate with disk warps because a variation of the PA causes a similar symmetric distortion in the isovelocity contours, as clearly visible in the innermost parts of the velocity map in Fig.\,\ref{fig:maps}. The first run of \textsc{$^{\rm 3D}$Barolo} does not provide strong evidence for a warped disk in the inner regions, considering both PA and inclination (see Fig.\,\ref{fig:BB}). All data points, indeed, are consistent with the mean PA and inclination within about 1$\sigma$ uncertainties. To determine the direction of the radial motions (inflow or outflow), one needs to know which side of the disk is nearest to the observer. The dust lanes visible in the HST image (Fig.\,\ref{fig:HST}, left panel) strongly suggest that the western side is the near one, so the disk is rotating in anti-clockwise direction and the radial motions are an inflow. Consequently, both the inner dusty spiral arms and the outer gaseous arms/tails are trailing, as is generally expected \citep[e.g.,][]{Sellwood2021}.

To further investigate non-circular motions, Figure \ref{fig:Maps} shows the observed, model, and residual velocity fields. The velocity fields are cropped beyond $R\simeq3.5''$ because we do not model the CO kinematics at larger radii due to the complex tail features. In general, the velocity residuals are below $\pm$30 \kms, indicating that there are no large-scale non-circular motions beyond the radial flow. The velocity residuals are higher than 30 \kms towards the Western edge of disk, where the S/N ratio of the CO line decreases and the kinematic disturbance of the tail may be important. There is also a compact yet significant velocity residual (below $-$30 \kms) near the galaxy center, which we discuss in detail below.

Figure \ref{fig:PV} shows position-velocity (PV) diagrams along the major and minor axis of the disk. Overall, the observed PV diagrams are well reproduced by our model. In particular, the minor-axis PV is not perfectly symmetric (as expected for purely circular orbits) but the CO flux density is enhanced and reduced in pairs of opposite quadrants, as defined by the disk center and systemic velocity. This asymmetry is well reproduced by our disk model, confirming the presence of radial motions. Both the major-axis and minor-axis PV diagrams show kinematically anomalous gas close to the center ($R\simeq0$) at velocities between $-100$ and $-200$ \kms, which cannot be reproduced by our disk model with rotation plus radial motions. This feature corresponds to the negative velocity residuals below $-$30 \kms\ in Figure\,\ref{fig:Maps} (right panel). One possibility is that this CO emission is probing the Keplerian rise of the rotation curve in the innermost (unresolved) parts of the molecular disk if the nuclear CO distribution is asymmetric between approaching and receding sides. Another possibility is that there are complex non-circular motion near the center, beyond a simple radial inflow, which may be due to stellar and AGN feedback. It is clear, however, that feedback must have a localized, small-scale effect ($\sim$100 pc) on the CO kinematics rather than a global, kpc-scale effect affecting the overall structure of the molecular gas disk.

\subsection{Mass models \& rotation-curve fits}\label{sec:mcmc}

In the previous section, we showed that the CO emission is well described by a rotating disk albeit there are significant non-circular motions in the inner parts, including a radial infall at $R<1''$. Since the gas kinematics remain dominated by regular rotation, it is sensible to use the observed rotation curve to study the inner mass distribution of Fairall\,49, assuming that the molecular gas is in dynamical equilibrium.

We build a mass model with four components: a stellar disk, a cold gas disk, a central SMBH, and a dark matter (DM) halo. The model has six parameters: disk inclination ($i$), stellar mass ($M_{\star}$), gas mass ($M_{\rm gas}$), SMBH mass ($M_{\bullet}$), halo mass ($M_{200}$), and halo concentration ($C_{200}$). The model parameters are determined using a MCMC method in a Bayesian context. We define the likelihood $\mathcal{L} = \exp({-0.5 \chi^2})$ with
\begin{equation}
 \chi^2 = \sum_{k}^{N} \frac{[V_{\rm rot}(R_k) - V_{\rm mod}(R_k)]^{2}}{\delta^{2}_{V_{\rm rot}} (R_k)},
\end{equation}
where $V_{\rm rot}$ is the observed rotation velocity at radius $R_k$, $\delta_{V_{\rm rot}}$ is the associated error, $V_{\rm mod}$ is the model rotation velocity, $k$ is the ring index, and $N$ is the total number of rings. \textsc{$^{\rm 3D}$Barolo} provide asymmetric errors ($\delta_{+}$, $\delta_{-}$) on $V_{\rm rot}$ corresponding to a variation of 5$\%$ of the residuals from the global minimum: we compute symmetric 1$\sigma$ errorbars as $\sqrt{\delta_{+}\delta_{-}}$.

Both $V_{\rm rot}$ and $\delta_{V_{\rm rot}}$ depend on the disk inclination $i$ and transform as $\sin(i_0)/\sin(i)$ where $i_0$ is the starting estimate. We assume a Gaussian prior on $i$ with a central value of 59$^{\circ}$ and a standard deviation of $\pm3^{\circ}$, as suggested by the tilted-ring fits with \textsc{$^{\rm 3D}$Barolo}.

The model velocity $V_{\rm mod}$ is obtained by summing in quadrature the velocity contribution of each mass component:
\begin{equation}\label{eq:Vmod}
 V_{\rm mod}^2 = \Upsilon_{\star}V_{\star}^2 + \Upsilon_{\rm gas}V_{\rm gas}^2 + \Upsilon_{\bullet}V_{\bullet}^2 + V_{\rm halo}^2(V_{200}, C_{200})
\end{equation}
where $\Upsilon_{\rm gas} = M_{\rm gas}/(10^9 M_{\odot})$ and $\Upsilon_{\bullet}= M_{\bullet}/(10^9 M_{\odot})$ are dimensionless factors introduced for numerical convenience, and $\Upsilon_{\star} = M_\star/L$ is the stellar mass-to-light ratio in solar units in the F606W band of HST/WFPC2 (similar to the $R$ band). In practice, $V_{\rm gas}$ and $V_{\bullet}$ are computed for an arbitrary mass of 10$^{9}$ M$_\odot$ and rescaled using $\Upsilon_{\rm gas}$ and $\Upsilon_\bullet$, respectively, so that the fitting parameters have values of the order of $\sim$1. Similarly, $V_{\star}$ is normalized to the total galaxy luminosity and rescaled via $\Upsilon_\star$.
Thus, the fitting parameters $\vec{\alpha}=\{M_\star, M_{\rm gas}, M_{\bullet}, M_{200}, C_{200}\}$ are mapped into $\vec{\beta} = \{ \Upsilon_{\star}, \Upsilon_{\rm gas}, \Upsilon_{\bullet}, V_{200}, C_{200} \}$.

The stellar gravitational contribution ($V_\star$) is computed using the techniques described in Sect.\,\ref{sec:stellarpot}. In particular, we stress that the innermost HSB component of the galaxy most likely does not trace the underlying stellar mass distribution because it is dominated by AGN and/or starburst emission. In our baseline mass model, this inner HSB component is subtracted (see inset in Fig.\,\ref{fig:sbp}) and neglected in the calculation of $V_{\star}$. Essentially it is assigned $\Upsilon_\star=0$. Different mass models in which the HSB component is treated as a sphere with a distinct $\Upsilon_{\rm HSB}$ are discussed in Sect.\,\ref{sec:mass}, while different ways of subtracting the HSB component are discussed in Appendix\,\ref{sec:sys}. These alternative models are used to estimate systematic uncertainties. We adopt a Gaussian prior on $\Upsilon_\star$ with a central value of 1.5 and a standard deviation of $\pm$1, as motivated by stellar population synthesis models of star-forming galaxies \citep{Schombert2019, Schombert2022}. In fitting rotation curves, the use of empirically motivated Gaussian priors is preferable to flat priors to break parameter degeneracies \citep{Li2019, Li2020} and avoid unphysical results \citep{Li2021}.

\begin{table}
\caption{Results from the rotation-curve fit. Errors correspond to the 68\% confidence interval of the marginalized 1-dimensional posterior distributions}
\label{tab:MCMC}
\centering
\begin{tabular}{l l}
\hline
Parameter                   & Best-fit value\\
\hline
$i$ ($^{\circ}$)            & 58.3$_{-3.0}^{+3.0}$ \\
$M_{\star}$ (10$^{9}$) M$_\odot$ & 21.5$_{-1.4}^{+1.7}$ \\
$M_{\bullet}$ (10$^{9}$ M$_\odot$) & 0.16$_{-0.04}^{+0.04}$ \\
$M_{\rm gas}$ (10$^{9}$ M$_\odot$) & 0.34$_{-0.29}^{+0.26}$ \\
$M_{200}$ (10$^{11}$ M$_\odot$) & 6.1$_{-4.4}^{+3.8}$ \\
$C_{200}$  & 6.8$_{-1.7}^{+1.7}$ \\
\hline
\end{tabular}
\end{table}

The gravitational contribution of the gas disk ($V_{\rm gas}$) is computed in a similar fashion as for the stellar disk, using the observed CO$(2-1)$ surface brightness profile in the radial direction (Fig.\,\ref{fig:BB}, top-right panel) and assuming an exponential density profile in the vertical direction with $h_{\rm z} =100$ pc (as in Sect.\,\ref{sec:Barolo}). We neglect the atomic gas contribution because \hi\ observations are not available for Fairall\,49. Atomic gas generally dominates the global gas budget in galaxies \citep[e.g.,][]{McGaugh2020}, but is subdominant in the innermost regions of massive galaxies \citep[see][]{Frank2016}, where the \hi\ surface density profiles display central depressions \citep{Martinsson2016} and the H$_2$ surface densities can reach high values of 100$-$1000 M$_\odot$ pc$^{-2}$. At any rate, as we will show, the gas contribution is subdominant with respect to the stellar one. We adopt a broad (nearly uninformative) log-normal prior centered at $\log(\Upsilon_{\rm gas}) = 0$ with a standard deviation of 0.5 dex.

The gravitational contribution of the SMBH $(V_{\bullet})$ is computed assuming a central point source with a mass of 10$^{9}$ M$_\odot$, which is scaled during the fit via $\Upsilon_{\bullet}$. We adopt a broad log-normal prior centered at $\log(\Upsilon_{\bullet}) = 0$ with a standard deviation of 0.5 dex.

The gravitational contribution of the DM halo $(V_{\rm DM})$ is computed assuming spherical symmetry and a Navarro-Frenk-White (NFW) density profile \citep{Navarro1996}. The observed rotation curve extends out to only 1.5 kpc, so the DM contribution is strongly subdominant and nearly unconstrained at these radii, but it is included for completeness. The halo mass $M_{200}$ and halo velocity $V_{200}$ are measured at $R_{200}$, the radius within which the mass volume density equals 200 times the critical density of the Universe. These two quantities are related by $V_{200} = [10\,G\,H_0\,M_{200}]^{1/3}$. The halo concentration is defined as $C_{200} = R_{200}/R_h$, where $R_h$ is the characteristic scale-length of the NFW profile. To constrain the halo parameters, we follow \citet{Li2020} and impose two $\Lambda$CDM scaling relations: (i) the $M_\star-M_{200}$ relation from abundance$-$matching techniques \citep{Moster2013}, and (ii) the $M_{200}-C_{200}$ relation from cosmological simulations \citep{Dutton2014}. In practice, for a given $M_\star$ during the MCMC simulation, $M_{200}$ and $C_{200}$ are assigned according to log-normal distributions centered on the $\log(M_\star)-\log(M_{200})$ and $\log(M_{200})-\log(C_{200})$ relations, respectively, assuming standard deviations of 0.15 dex in $\log(M_{200})$ \citep{Moster2013} and 0.11 dex in $\log(C_{200})$ \citep{Dutton2014}.

The posterior probability distributions of the parameters are mapped using \texttt{emcee} \citep{Foreman2013}. The MCMC chains are initialized with 200 walkers. Their starting positions are randomly assigned within sensible ranges for the fit parameters: $56^{\circ}<i<62^{\circ}$, $0.5 < \Upsilon_{\star} < 2.5$, $-1 < \log(\Upsilon_{\rm gas}) < 1$, $-1 < \log(\Upsilon_{\rm BH}) < 1$, $1.0 < \log(V_{\rm 200}) < 2.5$, and $0.6 < C_{200} < 1.2$. The starting positions of the walkers have virtually no effect on the final results but help to ensure fast convergence. We run 1000 burn-in iterations, then the sampler is run for another 2000 iterations. The \texttt{emcee} parameter $a$, which controls the size of the stretch move, is set equal to 2. This gives acceptance fractions around 50\%.

The MCMC results are summarized in Table \ref{tab:MCMC}. The errors correspond to the 68\% confidence interval of the marginalized 1-dimensional posterior probability distributions. The full posterior distributions of the fitting parameters are shown in Appendix\,\ref{sec:app}.

\subsection{Supermassive black hole mass}\label{sec:mass}

\begin{figure}
	\includegraphics[width=0.48\textwidth]{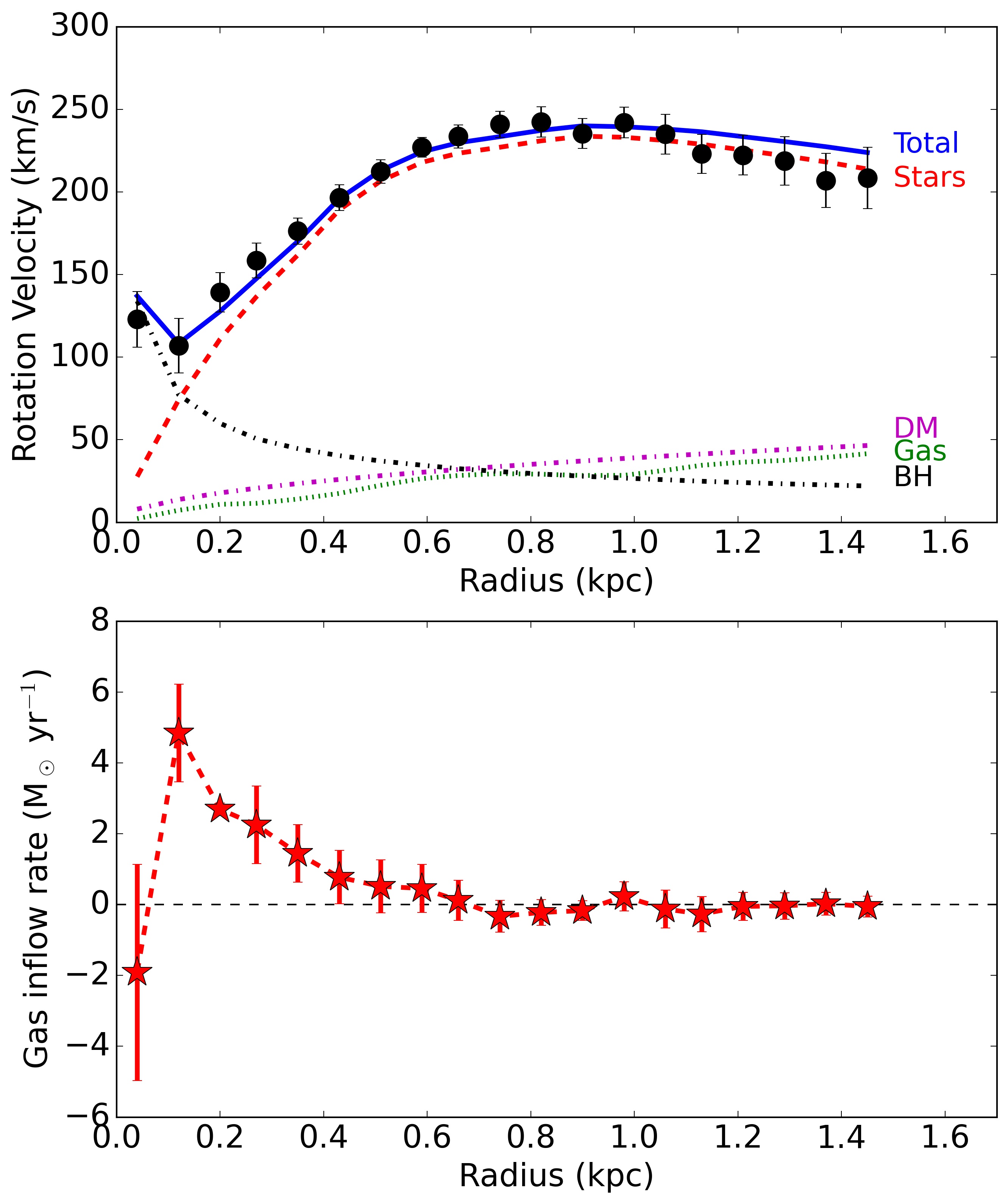}
    \caption{\textit{Top panel:} The observed rotation curve (black dots with errorbars) is fitted with a mass model (blue solid line) that considers the gravitational contributions of stars (red dashed line), gas (green dotted line), black hole (black dash-dotted line) and DM halo (purple dash-dotted line). \textit{Bottom panel:} gas inflow rate considering the molecular gas mass from the mass model. The error bars comprise the uncertainties on $V_{\rm rad}$ only; the uncertainties on $M_{\rm gas}$ have a systematic effect, shifting the inner peak up and down (see Eq.\,\ref{eq:inflow}).}
    \label{fig:massmodel}
\end{figure}

Figure\,\ref{fig:massmodel} (top panel) shows our baseline mass model. To explain the inner parts of the rotation curve, the mass model requires a central SMBH with a mass of $1.6 \pm 0.4 (\mathrm{rnd}) \pm 0.8 (\mathrm{sys}) \times 10^{8}$ $M_\odot$, where the random error corresponds to the 68\% confidence interval of the posterior probability distribution and the systematic error accounts for the subtraction of the AGN and starburst emission (see Appendix\,\ref{sec:sys}). For the baseline mass model in Figure\,\ref{fig:massmodel}, the SOI of the SMBH occurs at $R\simeq150$ pc (within which $V_\bullet > V_\star$), so it is well resolved by the angular resolution of the ALMA data ($\sim$80 pc).

Alternatively, we explored mass models in which the HSB component (red circles in Fig.\,\ref{fig:sbp}) is modeled as a sphere with a separate $\Upsilon_{\rm HSB}$, using the classic equation of \citet{Kent1986}. This spherical HSB component extends out to $\sim$200 pc and gives a velocity contribution that peaks at $R\simeq120$ pc, so we still need a central point mass to reproduce the declining part of the rotation curve at smaller radii. Assuming starburst-like values of $\Upsilon_{\rm HSB} \simeq 0.1-0.2$ \citep[e.g.,][]{McGaugh2014}, the resulting SMBH mass agrees with that of our baseline model within 1$\sigma$. Higher values of $\Upsilon_{\rm HSB}$ appears unphysical because (1) they are not expected from standard models of starburst and/or AGN emission, and (2) the shape of the rotation curve will not be reproduced. 

The only way to remove the need of a SMBH is to assign a substantial fraction ($\sim$15$\%$) of the luminosity of the HSB component to a point-like (unresolved) nuclear star cluster with $\Upsilon_\star\simeq1$, or equivalently a larger light fraction with a smaller $\Upsilon_\star$. This possibility seems unlikely because the HSB component is resolved by the HST resolution (Sect.\,\ref{sec:photo}) and display a Seyfert-1h optical spectra (Sect.\,\ref{sec:target}), so its innermost (unresolved) part must be dominated by the AGN. 

In addition to the modeling of the stellar gravitational potential, one may wonder whether a different modeling of the gas kinematics may remove the need for a SMBH. For example, the inner (unresolved) nuclear disk may have a different orientation of the kpc-scale gas disk. If there was a change in the inner PA, the ``true'' rotation velocities would be higher than derived here because our disk model would not consider the ``true'' kinematic major axis, so the evidence for a SMBH would actually increase. If there was a change in the inner inclination, the rotation velocities would substantially increase for an face-on warp (thus increasing the BH mass), but only moderately decrease for an edge-on warp. For example, an extreme change in inclination of 30$^{\circ}$ (from 58$^{\circ}$ to 88$^{\circ}$) would change the rotation velocities by a mere 18$\%$ (comparable to the random errors on the velocity points). The presence of a SMBH is confirmed in Sect.\,\ref{sec:KinMS} using the same forward-modeling technique as in previous WISDOM papers \citep[e.g.][]{Davis2020}, which fully takes into account the possible degeneracies between geometric parameters and SMBH mass.

\subsection{Gas mass \& gas inflow rate}

The molecular gas mass and the DM halo mass are less well determined than the SMBH mass because these two components are strongly subdominant at each radius. Thus, dynamical measurements of molecular gas masses remain challenging even when high-quality CO data are available. Our dynamical measurement of the H$_2$ mass, however, has important implications on the CO-to-H$_2$ conversion factor ($X_{\rm CO}$) in Fairall\,49, as we discuss in Sect.\,\ref{sec:XCO}.

Figure\,\ref{fig:massmodel} (bottom panel) shows the gas inflow rate ($\dot{M}_{\rm gas}$) at each radius $R$. To a first order approximation, $\dot{M}_{\rm gas}$ can be estimated as
\begin{equation}\label{eq:inflow}
    \dot{M}_{\rm gas}(R) = M_{\rm gas}(R) \dfrac{V_{\rm rad}(R)}{R},
\end{equation}
where $V_{\rm rad}$ is the radial velocity (defined to be positive for inflow, negative for outflow) from the tilted-ring analysis (Sect.\,\ref{sec:Barolo}) and $M_{\rm gas}$ is the molecular gas mass within each ring according to the scaling factor $\Upsilon_{\rm gas}$ from the MCMC fit. Strictly speaking, Eq.\,\ref{eq:inflow} provides an upper limit on $\dot{M_{H_2}}$ because we cannot tell how much gas in each ring is directly involved in the radial flow. The true value of $\dot{M}_{\rm gas}$ depends on angular momentum transport. For example, various studies \citep{GarciaBurillo2005, Haan2009, Querejeta2016} have estimated $\dot{M}_{\rm gas}$ in galaxy disks based on estimates of gravity torques on the gas and the resulting angular-momentum transport rates. They found that signs of radial motions from gas kinematics only loosely correlate with signs of average torques as a function of radius, so any value of $\dot{M}_{\rm gas}$ must be taken with a grain of salt. Moreover, in the next section, we will show that the forward modeling with \textsc{KinMS} provide significantly lower values of $V_{\rm rad}$ ($\sim$15 \kms rather than $\sim$75 \kms), so the gas inflow rates in Fig.\,\ref{fig:massmodel} may be up to a factor 5 smaller. The implications of this measurement on the feeding and growth of the SMBH are discussed in Sect.\,\ref{sec:baryonCycle}.

\begin{figure*}
	\includegraphics[width=\textwidth]{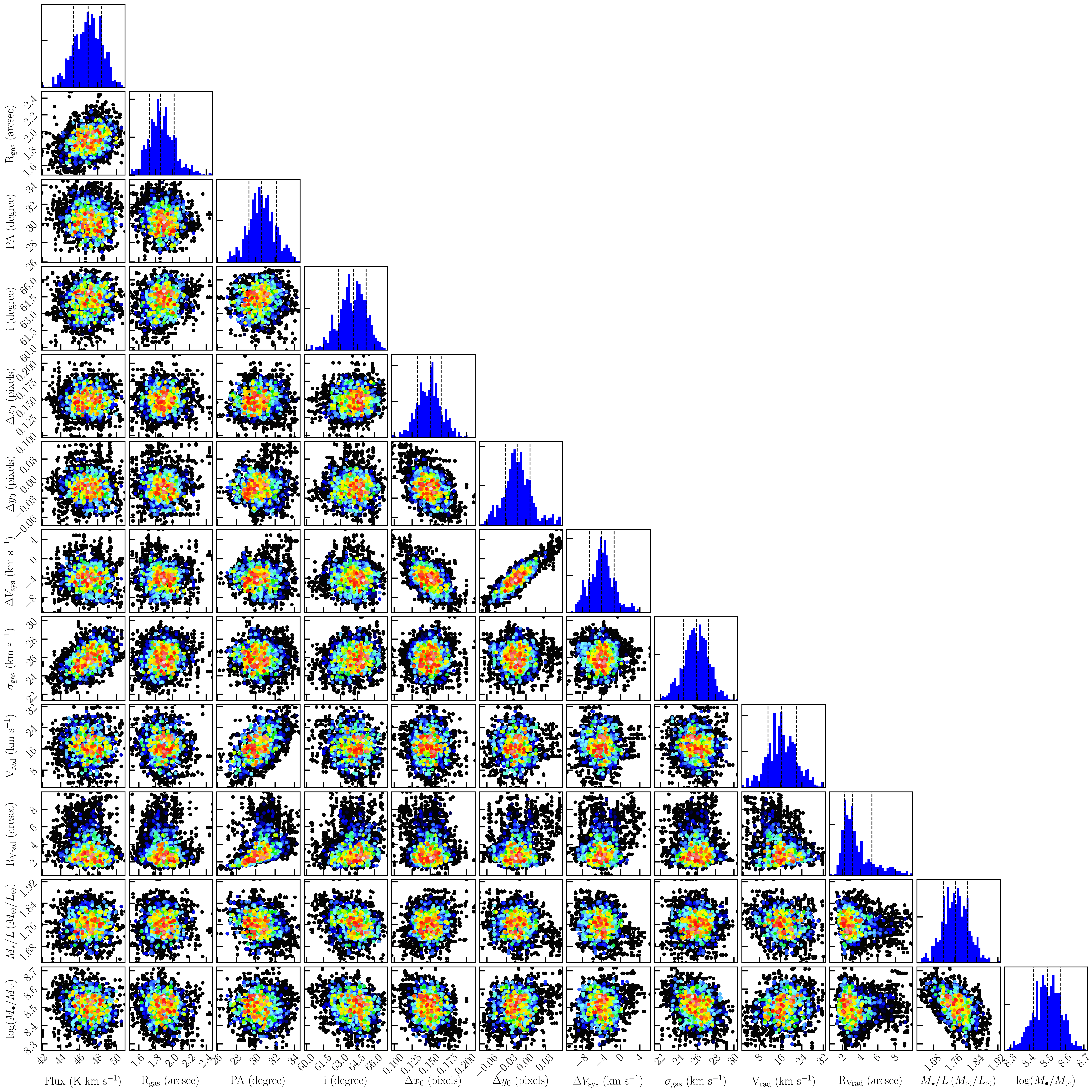}
    \caption{Posterior probability distributions of fitting parameters from the forward modeling with \textsc{KinMS}. In the marginalized 1D distributions, the dashed lines correspond to the 0.16, 0.50, and 0.84 quantiles (1$\sigma$ confidence intervals). In the 2D distributions, the rainbow colours indicate the maximum likelihood for a given combination of two parameters, with white corresponding to the maximum likelihood and dark blue to a 3$\sigma$ confidence level.}
    \label{fig:kinMS}
\end{figure*}
\subsection{Forward modeling}\label{sec:KinMS}

We use \textsc{KinMS} \citep{KinMS} to fit a mass model directly to the CO datacube. While \textsc{$^{\rm 3D}$Barolo} fits a kinematic model on a ring by ring basis with no assumptions on the mass distribution, \textsc{KinMS} allows fitting a global dynamical model to the datacube by specifying the gravitational potential at each radius. In this fashion, we can investigate possible degeneracies between dynamical parameters (such as $M_\star$ and $M_{\bullet}$) and kinematic ones (such as center, PA, and $i$) using an MCMC approach. This advantage comes at the expense of computation costs which are significantly higher for \textsc{KinMS} with respect to \textsc{$^{\rm 3D}$Barolo}. In analogy with previous WISDOM analyses, we use \textsc{KinMS} to fit only the inner parts of the CO disk ($R\simeq1"$). This is appropriate because we aim to (1) obtain an independent determination of the SMBH mass, whose dynamical influence is negligible at large radii, and (2) confirm (or not) the existence of radial motions, which occurs at $R<1"$ according to the \textsc{$^{\rm 3D}$Barolo} fits.

Our \textsc{KinMS} model has twelve free parameters. Two parameters ($F$ and $R_{\rm gas}$) describe the CO surface brightness distribution: $F$ is the CO flux within the fitted region of the cube, while $R_{\rm gas}$ is the CO scale-length adopting an exponential profile $\exp{(-R/R_{\rm gas})}$. Five parameters describe the orientation of the CO disk on the sky: PA and $i$ have the same meaning as in \textsc{$^{\rm 3D}$Barolo} fits (Sect. \ref{sec:Barolo}), while $\Delta x_0$, $\Delta y_0$, and $\Delta_{V_{\rm sys}}$ describe possible offsets with respect to the assumed dynamical center and systemic velocity. Another two parameters ($V_{\rm rad, 0}$ and $R_{\rm V_{rad}}$) are used to model radial motions, assuming a linear trend with radius motivated by Fig.\,\ref{fig:BB}: $V_{\rm rad, 0}(1 -R/R_{\rm V_{rad}})$. The last two free parameters ($\Upsilon_\star$ and $M_\bullet$) are related to the inner gravitational potential. We use Eq.\,\ref{eq:Vmod} to describe the circular velocity of a test particle in the disk, fitting for $\Upsilon_\star$ and $\log(M_\bullet) = \log(\Upsilon_\bullet)+9$ in analogy to previous WISDOM analyses. At $R<1"$ the gravitational contributions of gas and DM are negligible (see Fig.\,\ref{fig:maps}) but we include them for completeness fixing $\Upsilon_{\rm gas}$, $V_{200}$, and $C_{200}$ to the best-fit values found in Sect.\,\ref{sec:mass} from the full rotation curve.

\begin{table}\label{tab:kinMS}
\caption{Results from the forward modeling with \textsc{KinMS}}
\centering
\begin{tabular}{l c c}
\hline
Parameter                & Boxcar Prior Range & Best Fit Value \\
\hline
$F$ (K \kms)             & 10 to 100  & $46.9\pm1.5$\\
$R_{\rm gas}$ ($''$)     & 1.0 to 4.0 & $1.9^{+0.16}_{-0.13}$ \\
PA ($^{\circ}$)          & 30 to 60   & $30.6^{+1.6}_{-1.3}$ \\
$i$ ($^{\circ}$)         & 40 to 80   & $64.1\pm{1.2}$\\
$\Delta x_0$ (pixels)    & $-1$ to $+1$ & $0.15^{+0.01}_{-0.02}$\\
$\Delta y_0$ (pixels)    & $-1$ to $+1$ & $-0.01\pm0.02$\\
$\Delta V_{\rm sys}$ (\kms) & $-50$ to $+50$ & $-4.0\pm2.6$\\
$\sigma_{\rm gas}$ (\kms) & 1 to 100 & $25.9\pm1.3$\\
$V_{\rm rad, 0}$ (\kms) & 0 to 100 & $16.2^{+5.6}_{-5.0}$ \\
$R_{V_{\rm rad}}$ ($''$) & 0.1 to 10 & $3.1^{+2.2}_{-0.9}$\\
$\Upsilon_{\star}$       & 0.1 to 10 & $1.76\pm0.05$\\
$\log(M_{\bullet}/M_\odot$) & 5 to 10 & $8.50\pm0.07$ \\
\hline
\end{tabular}
\end{table}

We set the MCMC fit assuming uniform priors with sensible boundaries (see Table\,\ref{tab:kinMS}) and run 30000 samples. Figure\,\ref{fig:kinMS} shows the resulting posterior probability distributions. They are roughly Gaussians and display clear peaks, confirming the detection of both a SMBH and radial motions. The geometric parameters (PA, $i$, $\Delta x_{0}$, $\Delta y_{0}$, $\Delta V_{\rm sys}$) are consistent within the uncertainties with those from \textsc{$^{\rm 3D}$Barolo}. Intriguingly, $\Delta y_{0}$ and $\Delta V_{\rm sys}$ are degenerate but this has no effect on our general conclusions because they are both consistent with zero shift from the initial values.

The posterior probability distribution of $R_{V_{\rm rad}}$ show a broad tail towards high values, which is probably due to the fact that we fit only the inner CO disk at $R<1''$. The amplitude of the radial motions ($V_{\rm rad, 0}\simeq16$ \kms) is significantly lower than found with \textsc{$^{\rm 3D}$Barolo}, whereas the mean gas velocity dispersion is higher ($\sigma_{\rm gas}\simeq26$ \kms). In our experience, when a rotating disk model cannot perfectly reproduce some features in the observed 3D cube (due to, e.g., non-circular motions), automatic fitting codes tend to increase the gas velocity dispersion to ``cover up'' such detailed features with a broad line profile. The observed broadening of the line profiles, indeed, is driven by a complex interplay between spectral resolution, intrinsic gas velocity dispersion, beam-smeared rotation, and beam-smeared non-circular motions. It is likely that \textsc{KinMS} is compensating a lower $V_{\rm rad}$ with a higher $\sigma_{\rm gas}$, or alternatively $^{\rm3D}$Barolo is compensating a higher $V_{\rm rad}$ with a lower $\sigma_{\rm gas}$.

Regarding dynamical parameters, \textsc{KinMS} gives $\Upsilon_\star=1.76\pm0.05$ and $M_\bullet = (3.2 \pm 0.5) \times 10^8$ M$_\odot$, while the rotation-curve fit (Sect.\,\ref{sec:mass}) gives $\Upsilon_\star=1.8\pm0.1$ and $M_\bullet = (1.6 \pm 0.4) \times 10^8$ M$_\odot$, so the stellar mass-to-light ratios are virtually identical while the SMBH masses are consistent within 2$\sigma$ (random errors). If we use \textsc{KinMS} to fit a smaller area of the CO disk ($R<0.5"$), the value of $\Upsilon_\star$ increases to 2.3 while the SMBH mass decreases to $1.3 \times 10^8$ M$_\odot$, which is consistent with that from the rotation-curve fit within less than 1$\sigma$. A \textsc{KinMS} fit of such a small region, however, cannot constrain the radial motions occurring out to $R\simeq1''$. In general, we find overall agreement between forward-modeling with \textsc{KinMS} and rotation-curve fits. Hereafter, we adopt $M_\bullet = 1.6 \times 10^8$ M$_\odot$ as our fiducial SMBH mass.

\section{Discussion}\label{sec:discussion}

In the previous Sections, we found that the molecular gas disk of Fairall\,49 is well-described by regular rotation plus a radial inflow at $R \lesssim 500$ pc. After having subtracted the AGN and starburst contributions to the light profile, we built mass models to measure the SMBH mass, the molecular gas mass, and the gas inflow rate. In the following, we discuss the implications of these results on the co-evolution between SMBH and host galaxy (Sec.\,\ref{sec:coevolution}), the CO-to-H$_2$ conversion factor (Sec.\,\ref{sec:XCO}), and the baryon cycle in AGNs (Sec.\,\ref{sec:baryonCycle}).

\subsection{Implications for the SMBH and host galaxy}\label{sec:coevolution}

We measured the SMBH mass of Fairall\,49 fitting a mass model either to the rotation curve (Sect.\,\ref{sec:mass}) or directly to the CO datacube (Sect.\,\ref{sec:KinMS}). These two different methods consistently point to a SMBH with a mass of $\sim$10$^{8}$ M$_\odot$. As far as we are aware of, this is the first direct estimate of $M_\bullet$ in Fairall\,49. Previous indirect estimates of $M_\bullet$ used empirical scaling relations. \citet{Iwasawa2016} found $2\times10^6$ M$_\odot$ from X-ray variability; this value is in strong disagreement with our value being 50 times smaller. \citet{Dasyra2011} found $7\times10^{8}$ M$_\odot$ using the velocity dispersion of narrow-line regions (NLR) in the near-IR ($\sigma_{\rm NLR}$), and $1.7\times10^{8}$ M$_\odot$ using an improved relation between $M_\bullet$, $\sigma_{\rm NLR}$, and the NLR luminosity. This latter value is in good agreement with our dynamical determination. 

\begin{figure}
	\includegraphics[width=0.47\textwidth]{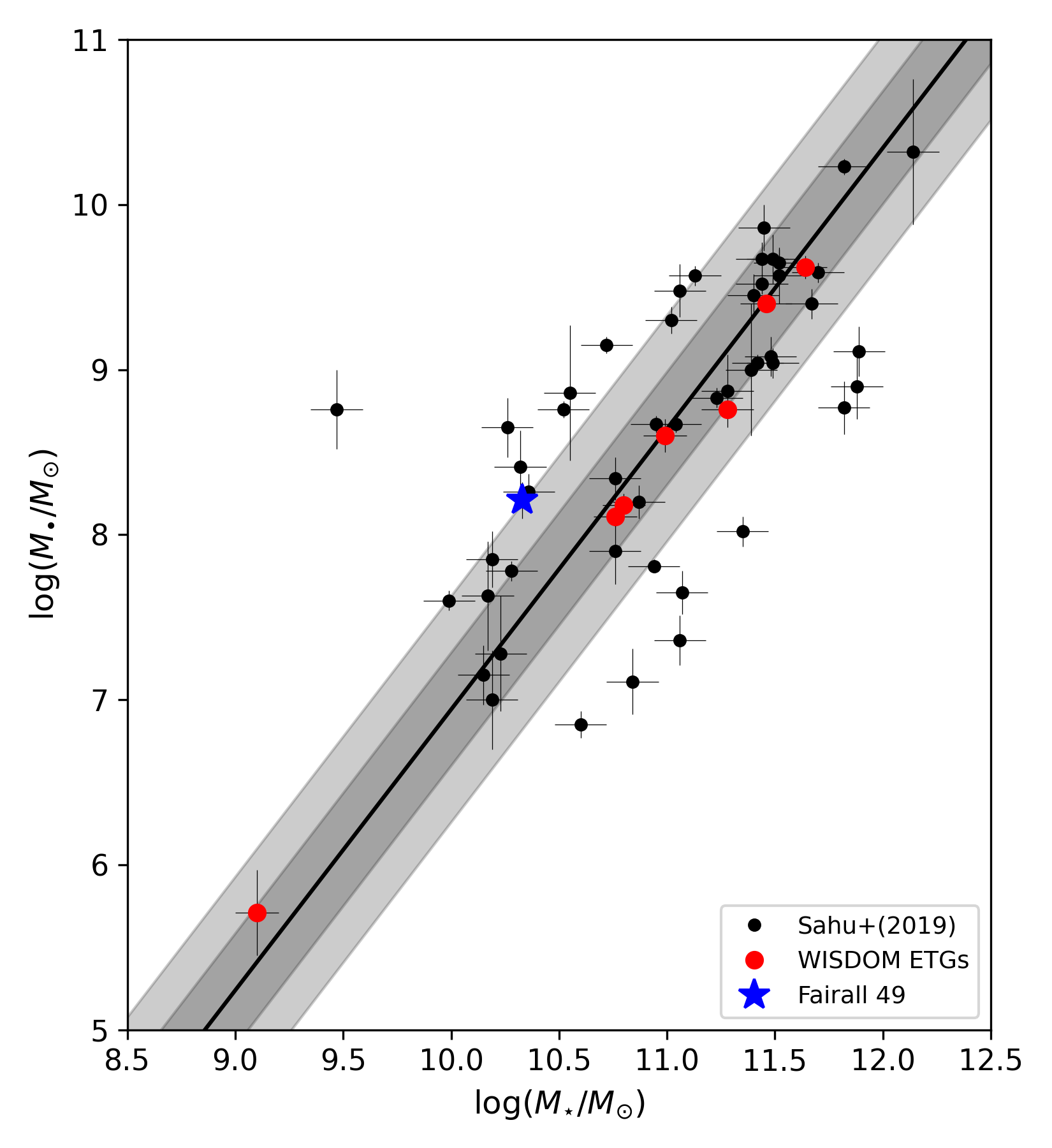}
    \caption{The location of Fairall\,49 (blue star) on the $M_\bullet-M_\star$ relation. Black and red dots show ETGs from \citet{Sahu2019} and the WISDOM papers, respectively. The solid line shows a fit using BayesLineFit \citep{Lelli2019}; the light and dark grey bands correspond to 1 and 2 times the orthogonal intrinsic scatter in the relation.}
    \label{fig:MBH}
\end{figure}
To investigate the location of Fairall\,49 on the $M_\bullet-M_\star$ relation, we consider the data from \citet{Sahu2019} for 83 ETGs and update them with six SMBH measurements from WISDOM \citep{WISDOM-1, WISDOM-2, WISDOM-3, WISDOM-4, WISDOM-5, WISDOM-7}. We also add the dwarf lenticular galaxy NGC\,404 \citep{Davis2020} as well as Fairall\,49. We fit the data using the BayesLineFit software \citep{Lelli2019}, which perform an MCMC analysis considering errors on both variables and a Gaussian model with constant intrinsic scatter along the orthogonal direction to the best-fit line. We find 
\begin{equation}
    \log(M_\bullet) = 1.75 \log(M_\star) - 10.6 
\end{equation}
with an orthogonal intrinsic scatter of 0.36 dex. The posterior probability distributions of slope, intercept, and intrinsic scatter are shown in Appendix\,\ref{sec:app}. Figure\,\ref{fig:MBH} shows that Fairall\,49 lies on the same $M_\bullet-M_\star$ relation as ETGs within 2$\sigma$ of its intrinsic scatter, giving further support to the reliability of our SMBH mass measurement.

The location of Fairall\,49 on the bulge mass$-$SMBH mass relation is unclear because the galaxy does not have a clear bulge (see Sect.\,\ref{sec:target}). The galaxy has an inner stellar disk resembling a late-type spiral, while its outer asymmetric stellar distribution may possibly resemble an elliptical galaxy. In fact the galaxy has been classified as E-S0 using low-resolution ground-based images that do not resolve its inner parts \citep{Makarov2014}. If we assume that Fairall\,49 is an elliptical galaxy with a central spiral disk, then the bulge mass essentially coincides with the stellar mass and we obtain a similar picture as Fig.\,\ref{fig:MBH}.

\subsection{Implications for the CO-to-H$_2$ conversion factor}\label{sec:XCO}

Fitting a mass model to the rotation curve of Fairall\,49, we find a molecular gas mass of about $3.4 \times10^8$ M$_\odot$. An independent estimate can be obtained from the CO($J=2\rightarrow1$) luminosity. If we assume that (1) the ratio $R_{21} = \mathrm{CO}(J=2\rightarrow1)/\mathrm{CO}(J=1\rightarrow0)$ is equal to unity as expected for thermalized gas, (2) the H$_2$-to-CO conversion factor\footnote{Following \citet{Bolatto2013} we define $X_{\rm CO, 20} = 10^{20}$ cm$^{-2}$ (K \kms)$^{-1}$ and adopt $X_{\rm CO, 20} = 2$ for the Milky Way for CO($J=1\rightarrow0$).} is $X_{\rm CO, 20}=2$ as for the Milky Way, and (3) the correction factor for Helium and heavier elements is 1.37 \citep[e.g.][]{McGaugh2020}, then we obtain a molecular gas mass of $\sim$10$^{10}$ M$_\odot$ that is $\sim$30 times higher than the dynamical estimate. This high value is surely ruled out because the resulting mass model would overshoot the observed rotation curve at large radii (Fig.\,\ref{fig:massmodel}). For example, even if we assume a decrease in $\Upsilon_\star$ by a factor of 2 in the outer parts (which seems unrealistic because $\Upsilon_\star$ should eventually decreases in the inner parts due to the on-going star formation), the resulting mass model would still overshoot the observed rotation curve by $\sim$50 \kms due to the substantial gas contribution. We conclude that at least one of the three assumptions above must be wrong. 

The corrections for Helium and heavier elements play a very minor role because realistic values range from 1.34 and 1.41 \citep{McGaugh2020}, so they are not going to affect the order of magnitude of our gas mass estimate. The values of $X_{\rm CO}$ and $R_{21}$, instead, may play a more important role. In starburst galaxies, $X_{\rm CO}$ could be lower than in normal galaxies like the Milky Way because the molecular gas have higher densities, temperatures, and velocity dispersions \citep{Bolatto2013}. For example, \citet{Zhu2003} finds $X_{\rm CO, 20}\simeq 0.2-0.4$ in the central regions of the Antennae galaxy merger, using multi-transition observations and gas excitation modeling. Similarly, \citet{Papadopoulos1999} find that $X_{\rm CO}\simeq 0.2-0.4$ in the starburst Seyfert-2 galaxy NGC\,1068, while \citet{Sliwa2012} find $X_{\rm CO}\simeq 0.2-0.6$ in the starburst Arp\,299. In addition to these in-depth analyses of individual systems, studies of large samples of LIRGs find that $X_{\rm CO, 20}$ ranges between 0.2 and 0.6 \citep{Yao2003, Papadopoulos2012}.

If we assume that the lower bound from these works ($X_{\rm CO, 20}=0.2$) applies to Fairall\,49, the luminosity-based gas mass would be $\sim$10$^9$ M$_\odot$ that is marginally consistent with the dynamical-based gas mass of $3.4^{+2.6}_{-2.9}\times10^8$ M$_\odot$ within 2.5$\sigma$. The residual tension may be alleviated with a radial variation of $X_{\rm CO}$: if $X_{\rm CO}$ is higher in the inner parts and lower in the outer regions, the shape of $V_{\rm gas}$ would be flatter so that $M_{\rm gas}$ can slightly increase without overshooting the observed rotation curve at large radii (see Fig.\,\ref{fig:massmodel}). In addition, there could be plausible variations in the line ratio $R_{21}$. Multi-transition CO observations have shown that $R_{21}$ can vary from $\sim$0.3 to $\sim$1.5 in typical spiral galaxies \citep{Leroy2009, Querejeta2021, denBrok2021}. Interestingly, in the Seyfert prototype NGC\,1068, the value of $R_{21}$ ranges from 1.6 to 2.5 \citep{Viti2014}. Thus, for a starburst-like value of $X_{\rm CO, 20}\simeq0.2$ and a Seyfert-like value of $R_{21}\simeq2.5$, the luminosity-based and dynamical-based gas masses of Fairall\,49 would be entirely consistent.

\subsection{Implications for the AGN baryon cycle}\label{sec:baryonCycle}

A basic result of this paper is that a LIRG with a central AGN can host a molecular disk in regular rotation. Similar disks of molecular gas have been observed in other nearby AGN-host galaxies \citep[e.g.,][]{Combes2019, Ruffa2019a, Ruffa2019} and also in AGN-host starburst galaxies at high $z$ \citep[e.g.,][]{Lelli2018, Lelli2021}. This suggests that AGN and starburst feedback must have a weak/moderate effect on the cold interstellar medium. Feedback may affect the local gas kinematics on scales of $\sim$100 pc, leading to local enhancements of the gas velocity dispersion and/or non-circular motions, but clearly does not destroy the global regularity of the gas disk on kpc scales.

In addition to the regular rotation, we detect a radial inflow motion in the central regions of Fairall\,49 ($R\lesssim$400 pc). The resulting gas inflow rate reaches a maximum of $\sim$5 M$_\odot$ yr$^{-1}$ for our baseline mass model. This value is in line with predictions of hydrodynamic simulations of SMBH fueling \citep{Alcazar2020}. We stress that this is an instantaneous gas inflow rate on scales of hundreds of pc: it is likely that the strength of radial motions vary at smaller spatial scales on short timescales, so that the instantaneous rate can change significantly. For example, the simulations of \citet{Alcazar2020} suggests that gas inflow rates can vary by more than 4 orders of magnitude on pc scales over time-scales as short as 1 Myr. Thus, it is difficult to draw any strong conclusions on the actual feeding of the SMBH near the event horizon.

We do not find strong evidence for a molecular gas outflow (see Sect.\,\ref{sec:Barolo}). Tentative evidence is represented by kinematically anomalous CO emission at $R\simeq0$ with excess velocities of $\sim$100 \kms with respect to the underlying rotation (see Fig.\,\ref{fig:PV}). We cannot tell whether this emission is due to a weak molecular outflow, complex non-circular motions within the disk, or the Keplerian rise of circular motions around the SMBH in a nuclear CO disk. In the first case, the gas outflow should be compact and collimated along the line of sight given the elongated shape in both PV diagrams (see Fig.\,\ref{fig:PV}), so the observed line-of-sight velocity would be comparable to the intrinsic outflow velocity. The velocities of the putative outflow, therefore, would be significantly smaller than the galaxy escape velocity\footnote{The galaxy escape velocity is calculated as $\sqrt{2|\Phi(r)|}$ along the direction perpendicular to the disk, where $\Phi(r)$ is the gravitational potential from a parametric model with a spherical NFW halo, an exponential stellar disk, an exponential gas disk, and a SMBH with the same masses as those in Table\,\ref{tab:MCMC}.} ($\sim$700 \kms\ at $R\lesssim 100$ pc) and the gas would eventually fall back on the system in a galactic-fountain fashion. Similar results have been found in other studies of individual AGN and starburst systems, which show that the velocities of molecular gas outflows are significantly smaller than the galaxy escape velocity \citep{Zanchettin2021, Fluetsch2019}. This includes the iconic starburst M82 \citep{Leroy2015}. The situation may be different in radio-loud AGNs, in which the radio jet can aid the development of fast outflows of both atomic and molecular gas with velocities of $\sim$600-800 \kms \citep{Oosterloo2017, Oosterloo2019}. From a more statistical perspective, stacking approaches using the Na\,{\small I}\,D absorption line shows that neutral-gas outflows are rare in the nearby Universe and have velocities below the galaxy escape velocity \citep{Concas2019}. Thus, Fairall\,49 seems to be the norm rather than the exception in terms of molecular gas outflows.

\section{Conclusions}\label{sec:summary}

In the context of the WISDOM project, we studied the starburst Seyfert galaxy Fairall\,49 using ALMA observations of the CO($2-1$) line with a spatial resolution of $\sim$80 pc. Our main results can be summarized as follows:
\begin{enumerate}
    \item The CO kinematics are well described by a rotating disk with a radial inflow at $R<400$ pc. The overall regularity of the molecular gas suggests weak feedback from both the starburst and AGN activity.
    \item The galaxy has a SMBH with a mass of
    $1.6\pm0.4\mathrm{(rnd)}\pm0.8 \mathrm{(sys)}\times 10^{8}$\,\Msun\ where the systematic error is driven by the subtraction of the AGN and starburst emission from the observed light profile (see Appendix\,\ref{sec:sys}). Our measurement of $M_\bullet$ is $\sim$50 times higher than previous estimates from X-ray variability.
    \item Fairall\,49 lies on the same $M_\bullet-M_\star$ relation as ETGs. The location of Fairall\,49 on the bulge mass$-$SMBH mass relation is unclear because the galaxy does not have a clear bulge.
    \item The molecular gas mass inferred from the rotation curve is 30 times smaller than expected for a Galactic $X_{\rm CO}$ and thermalized gas with $R_{21}=1$. This suggests low values of $X_{\rm CO}$ and high values of $R_{21}$ in line with other studies of starburst and Seyfert galaxies.
    \item For our fiducial estimate of the molecular gas mass, the gas inflow rate may be as high as $\sim$5 M$_\odot$ yr$^{-1}$. This is in agreement with the predictions of hydrodynamical simulations of AGN feeding.
\end{enumerate}
This work highlights the potential of using high-resolution ALMA data of nearby galaxies to probe, in addition to SMBH masses, the $X_{\rm CO}$ factor and nuclear gas flows. Future investigations will address these issues using the entire WISDOM sample.

\section*{Acknowledgements}

FL thanks James Schombert and Enrico Di Teodoro for their help in using \textsc{Archangel} and \textsc{$^{\rm 3D}$Barolo}, respectively. We also thank Marc Sarzi for useful comments. TGW acknowledges funding from the European Research Council (ERC) under the European Union’s Horizon 2020 research and innovation programme (grant agreement No. 694343). This paper makes use of the following ALMA data: ADS/JAO.ALMA\#2017.1.00904.S. ALMA is a partnership of ESO (representing its member states), NSF (USA) and NINS (Japan), together with NRC (Canada), MOST and ASIAA (Taiwan), and KASI (Republic of Korea), in cooperation with the Republic of Chile. The Joint ALMA Observatory is operated by ESO, AUI/NRAO and NAOJ. Based on observations made with the NASA/ESA Hubble Space Telescope, and obtained from the Hubble Legacy Archive, which is a collaboration between the Space Telescope Science Institute (STScI/NASA), the Space Telescope European Coordinating Facility (ST-ECF/ESA) and the Canadian Astronomy Data Centre (CADC/NRC/CSA).

\section*{Data Availability Statement}

The data underlying this article will be shared on reasonable request to the corresponding author.



\bibliographystyle{mnras}
\bibliography{bibliography} % if your bibtex file is called example.bib

\appendix

\section{Systematic uncertainties due to AGN and starburst light subtraction}\label{sec:sys}

\begin{figure*}
	\includegraphics[width=0.24\textwidth]{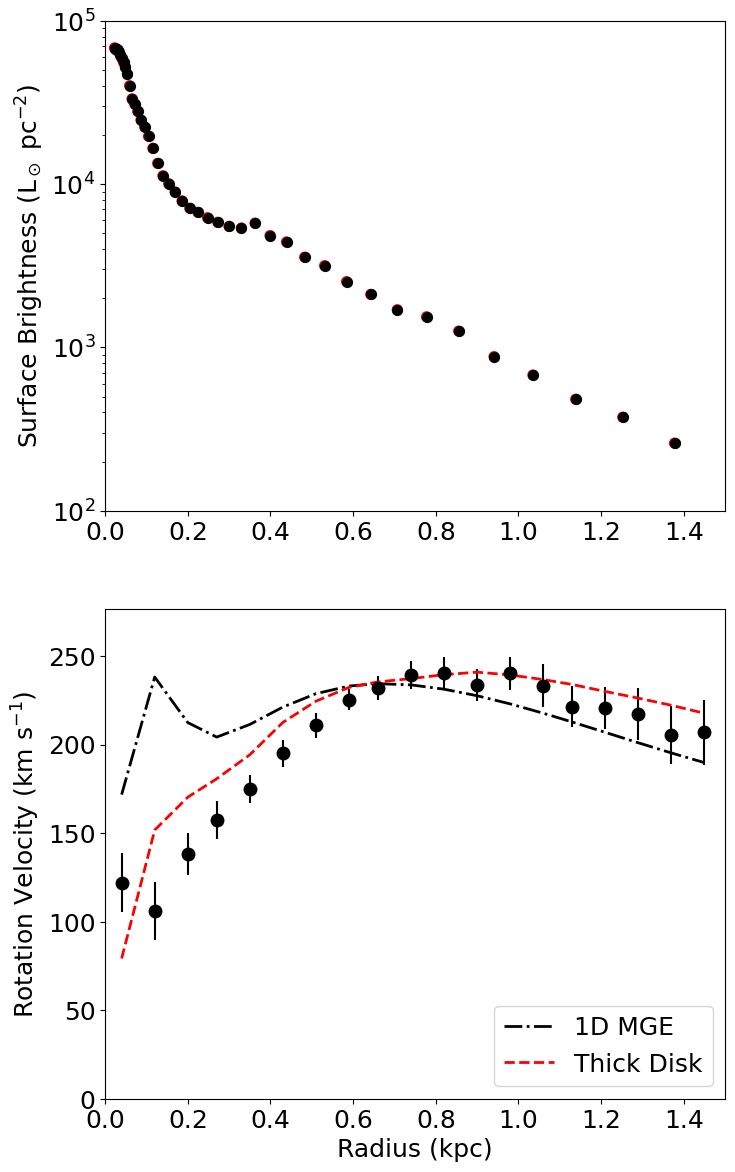}
	\includegraphics[width=0.24\textwidth]{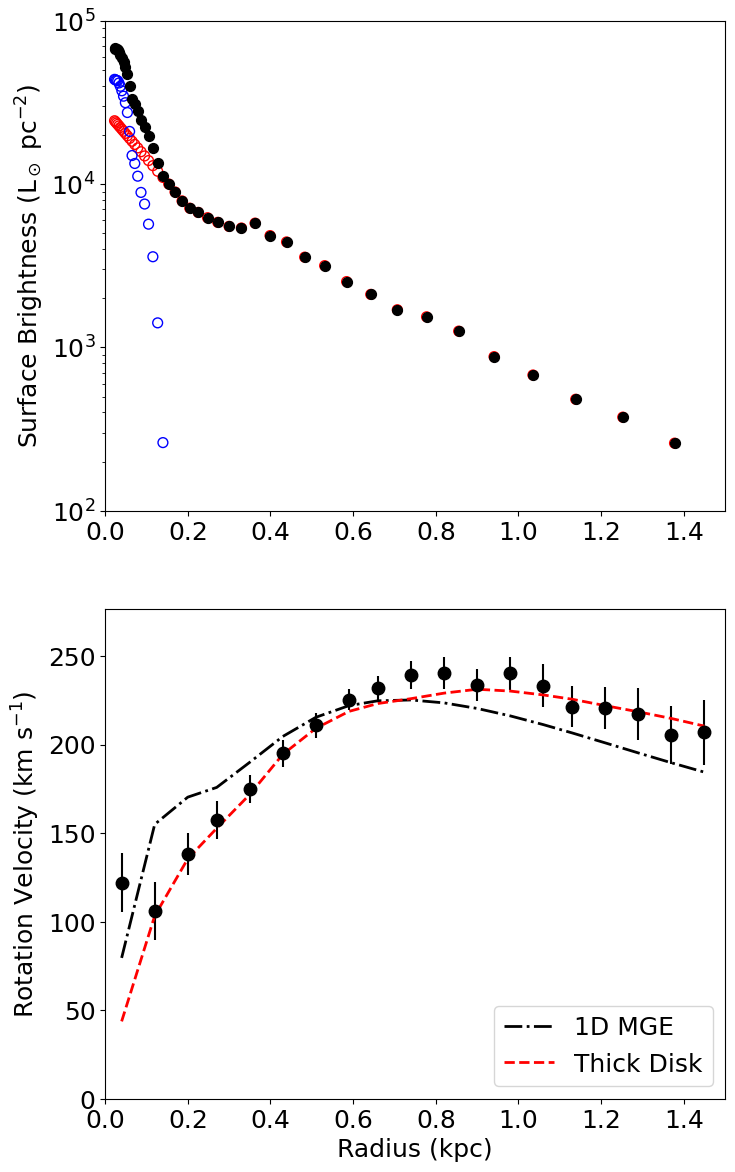}
	\includegraphics[width=0.24\textwidth]{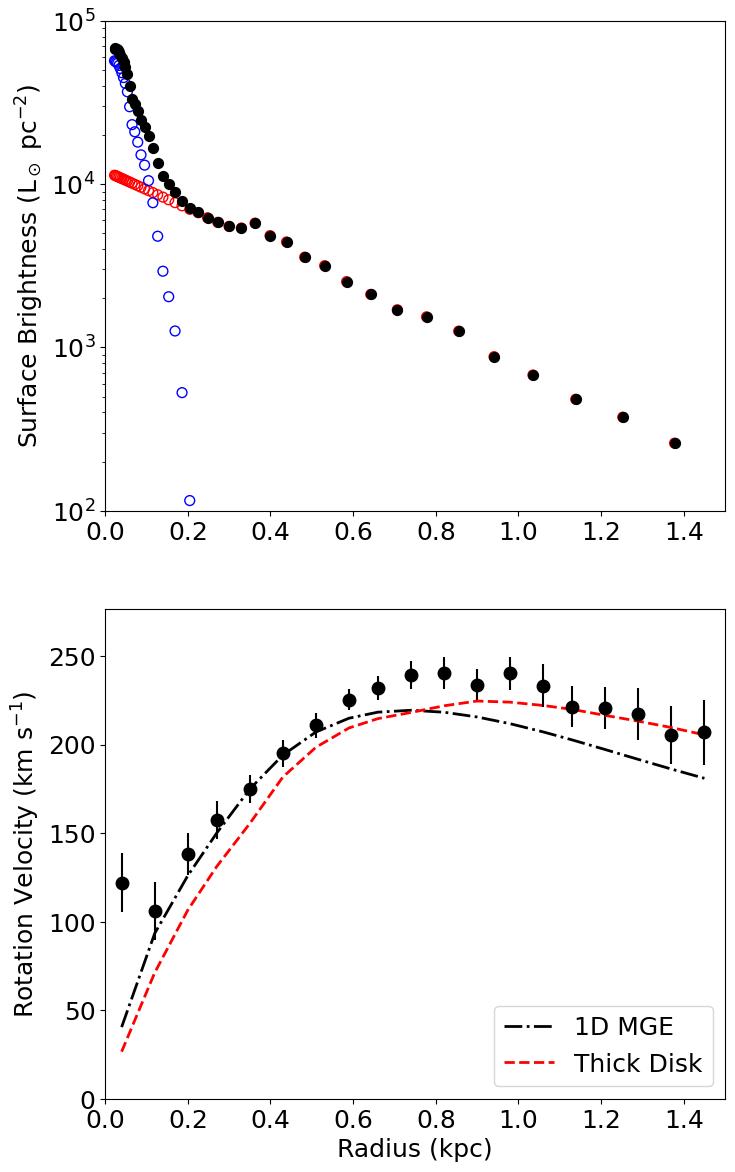}
	\includegraphics[width=0.24\textwidth]{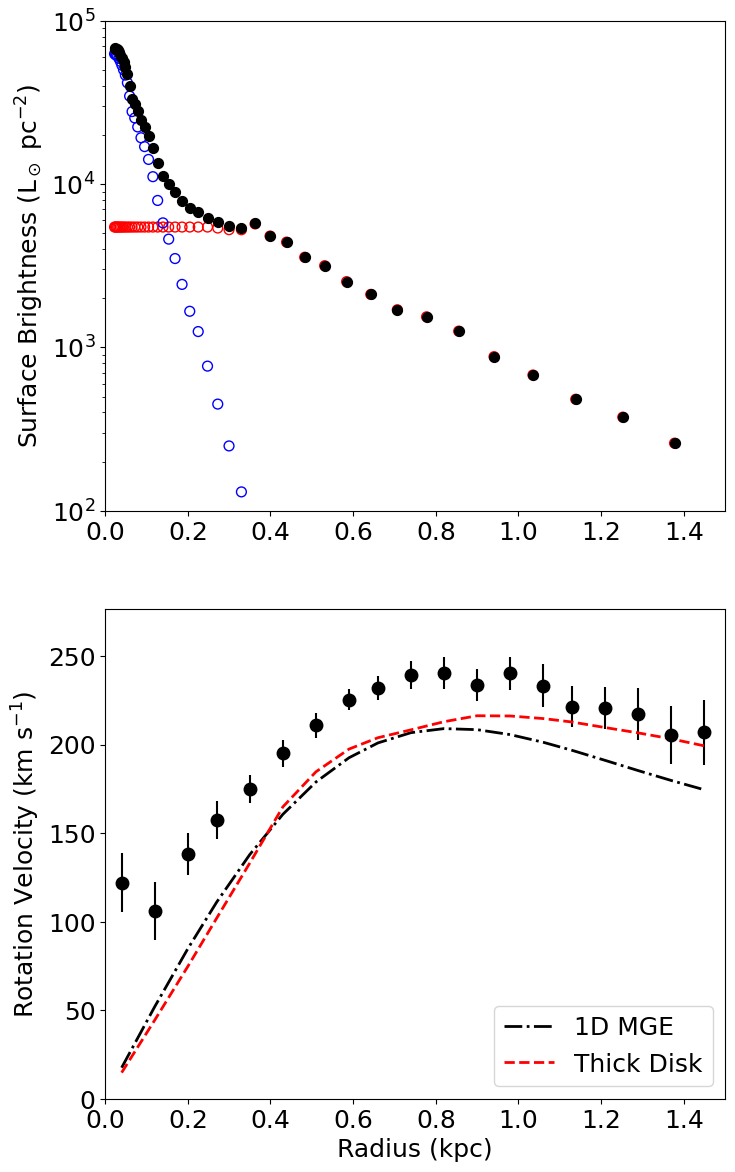}
    \caption{\textit{Top panels:} Four different ways of subtracting the inner HSB component (blue circles) from the observed surface brightness profile (black dots) by extrapolating an inner exponential profile (red circles). \textit{Bottom panels:} the observed rotation curve (black dots) is compared with the stellar velocity contributions obtained for the different subtraction models, using a 1D-MGE approach (black dot-dashed line) and a thick-disk approach (red dotted line). See text for details.}
    \label{fig:subtest}
\end{figure*}

\begin{figure*}
	\includegraphics[width=0.475\textwidth]{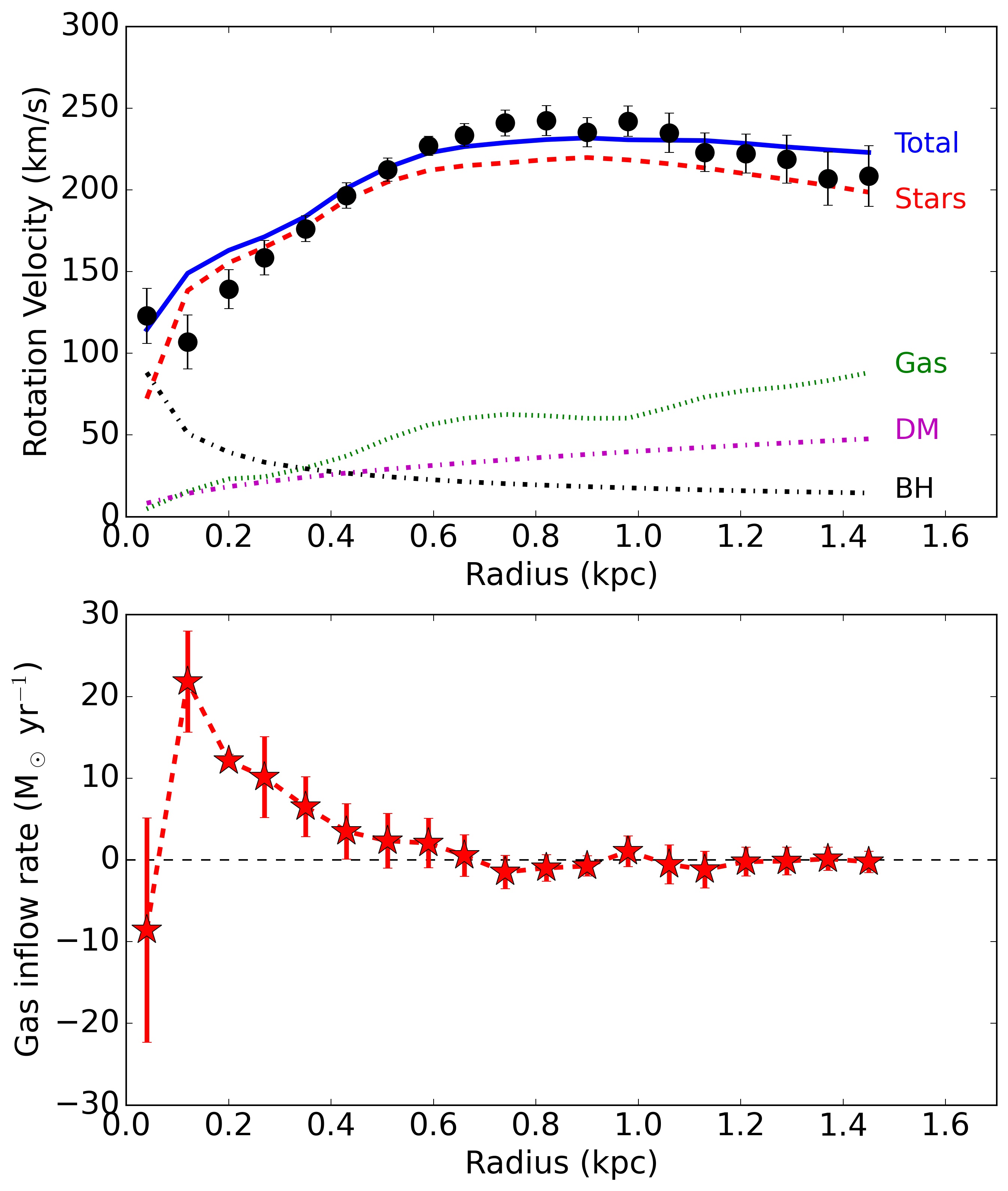}
	\includegraphics[width=0.475\textwidth]{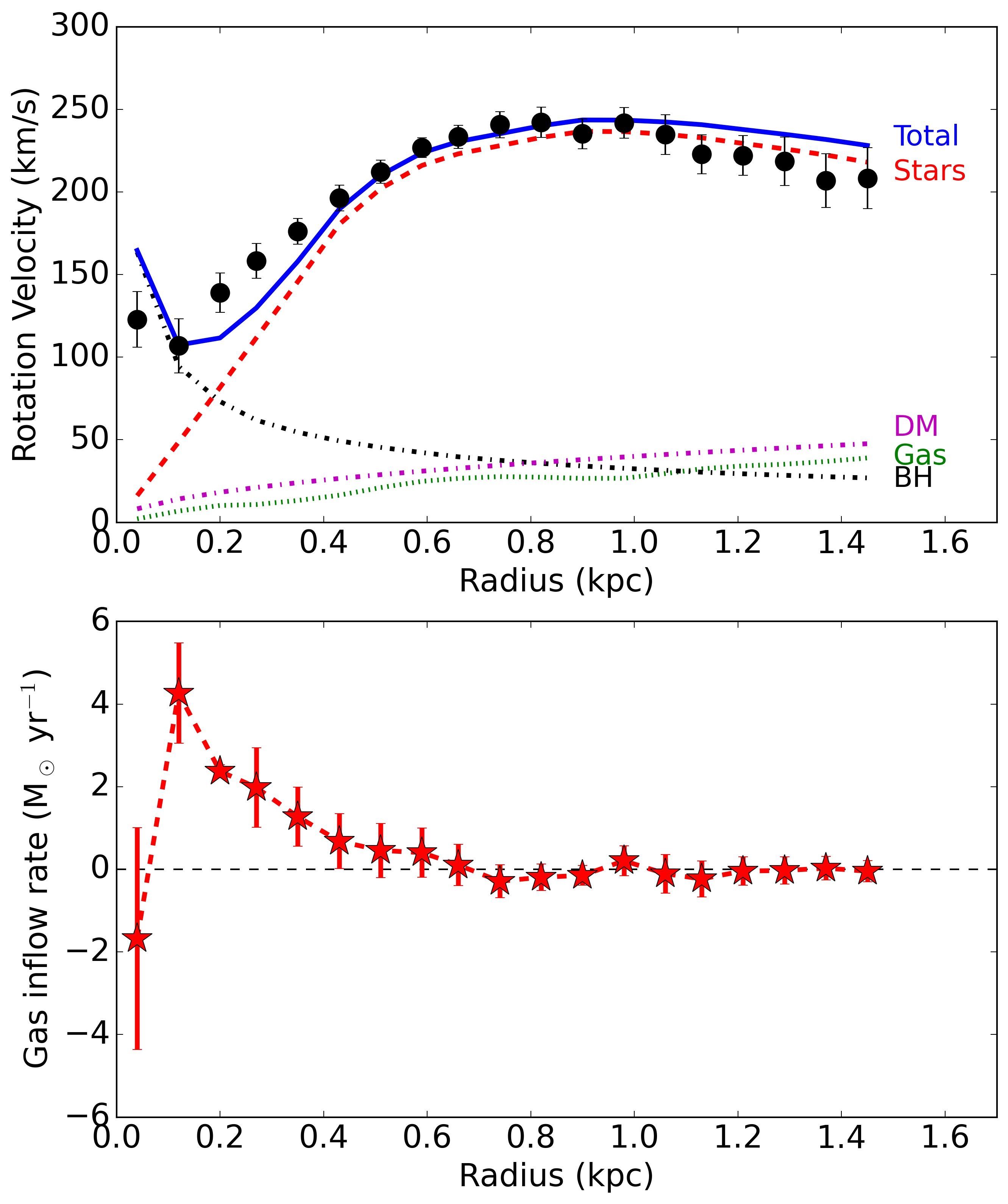}
    \caption{Same as Figure\,\ref{fig:massmodel} but calculating the stellar gravitational contribution with no HSB subtraction (left panel) and maximal HSB subtraction (right panel). Both mass models do not provide a good fit to the inner rotation curve and are probably unphysical, but can be used to estimate systematic uncertainties on the SMBH mass and gas inflow rate.}
    \label{fig:sys}
\end{figure*}

The top panels of Figure\,\ref{fig:subtest} illustrate different ways of subtracting the central HSB component from the surface brightness profile: (1) no subtraction, (2) a minimal subtraction of $\sim0.6 \times 10^9$ L$_\odot$ by extrapolating a steep exponential profile at $R<0.2$ kpc, (3) our baseline subtraction of $\sim10^9$ L$_\odot$ by extrapolating inwards the overall exponential slope at $0.2<R<1.5$ kpc, and (4) a maximal subtraction of $\sim1.6 \times 10^{9}$ L$_\odot$ by extrapolating a flat profile at $R<0.2$ kpc. The bottom panels of Figure\,\ref{fig:subtest} show the resulting circular velocity profiles adopting a nominal stellar mass-to-light ratio of 1.8 in the F606W band. The stellar gravitational contribution is computed using two different approaches: a 1D MGE fit \citep{Cappellari2002} and a thick disk \citep{Casertano1983}. See Sect.\,\ref{sec:stellarpot} for details.

In the case of (1) no subtraction and (2) minimal subtraction, there are significant differences between the 1D-MGE and thick-disk solutions. The 1D-MGE solutions cannot reproduce the observed rotation-curve shape, so they appear unphysical. Similar results are found using a 2D-MGE fit to the HST image (Fig.\,\ref{fig:HST}), even if we exclude the innermost Gaussian with full-width half-maximum comparable to the PSF of HST. On the other hand, the thick-disk solutions can reproduce the observed rotation curve but they appear unphysical too. In fact, if we project on the sky a thick disk with a luminosity profile given by models (1) or (2), the projected light distribution display a $\sim$90$^{\circ}$ twisting of the inner isophotes, which is not evident in the HST image. This occurs because the inner HSB component is effectively prolate (the assumed scale-height is larger than the effective scale-length within 200 pc), so it leads to isophotes oriented perpendicularly the projected major axis of the disk and provides limited gravitational force in the disk mid-plane. Nevertheless, these thick-disk solutions can be used to obtain conservative systematic uncertainties due to HSB subtraction and intrinsic 3D geometry.

In the case of baseline subtraction (3) and maximal subtraction (4), the 1D-MGE and thick-disk solutions give similar results, indicating that the intrinsic 3D geometry play a minor role as it is generally expected \citep[e.g.,][]{Noordermeer2008}. Most importantly, both solutions can reproduce the overall shape of the rotation curve. A similar stellar velocity contribution is found by removing the two innermost Gaussian from a 2D-MGE fit to the HST image.

If we use the four thick-disk models to estimate $V_\star$ in Eq.\,\ref{eq:Vmod} and fit the observed rotation curve, we obtain SMBH masses of 0.7, 1.2, 1.6, and 2.4 $\times$ 10$^8$ M$_\odot$, respectively. The fit results are shown in Figure\,\ref{fig:sys} for the most extreme models (1) and (4). Both mass models do not provide a good fit to the observed rotation curve, missing its inner shape and overestimating the outer rotation velocities. Both mass models are probably unphysical, but we can consider them as 1$\sigma$ systematic deviations from our baseline mass mass. Then, the systematic uncertainty on the fiducial SMBH mass of $1.6 \times 10^8$ M$_\odot$ is of the order of $\pm0.8$.

In addition, Figure\,\ref{fig:sys} shows that the mass model with no HSB subtraction gives a much higher gas inflow rate because the MCMC fit decreases the value of $\Upsilon_\star$ to fit the inner rotation-curve points while increasing that of $\Upsilon_{\rm gas}$ to fit the outer parts. Thus, the gas mass and the gas inflow rate from this model are substantially higher than those from the other three. As we already pointed out, the thick-disk model with no light subtraction is probably unphysical, so it should be taken with a grain of salt. The other mass models return a maximum gas inflow rate between $\sim$4 and $\sim$6 M$_\odot$ yr$^{-1}$. Thus, the main source of systematic uncertainty on the gas inflow rate is not the subtraction of the HSB component, but rather the values of the radial velocities that can vary by a factor of 5 when fitting the datacube with either \textsc{$^{\rm 3D}$Barolo} or \textsc{KinMS} (see Sect.\,\ref{sec:mass}).

\section{Posterior Probability Distributions}\label{sec:app}

\begin{figure*}
	\includegraphics[width=0.95\textwidth]{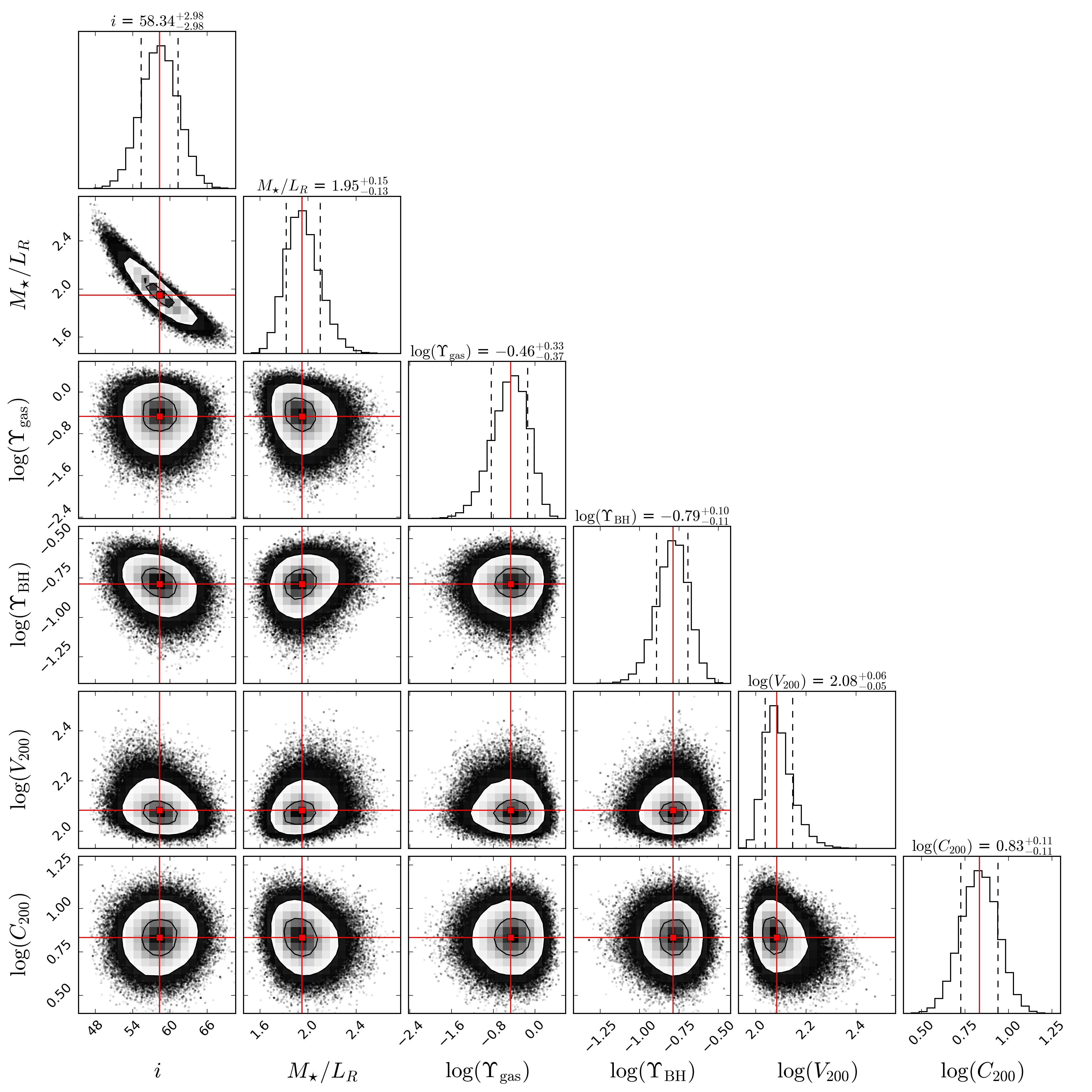}
    \caption{Posterior probability distributions from fitting the CO rotation curve with the baseline mass model (see Fig.\,\ref{fig:massmodel}).}
    \label{fig:corner2}
\end{figure*}

\begin{figure}
	\includegraphics[width=0.48\textwidth]{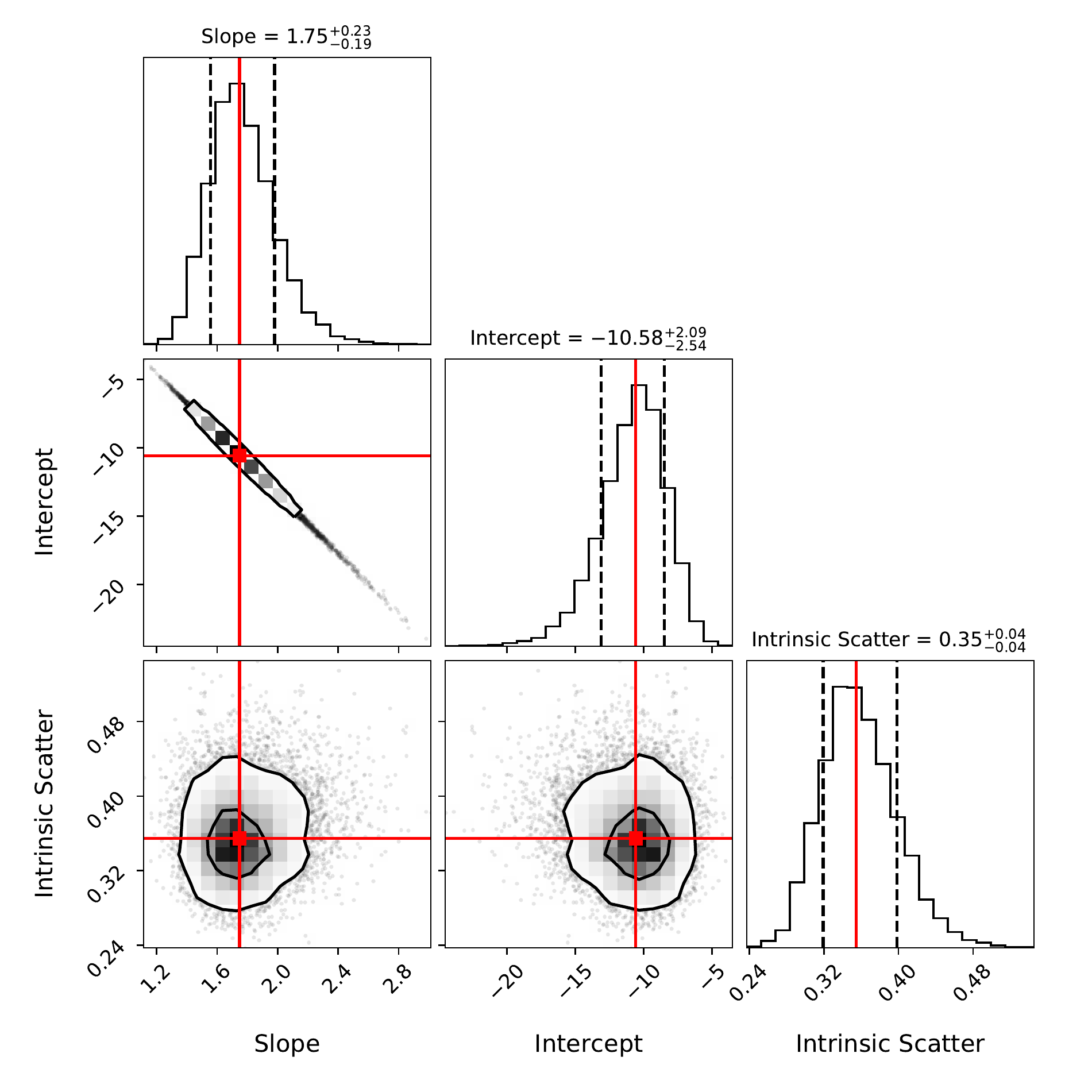}
    \caption{Posterior probability distributions from fitting the $M_\bullet-M_\star$ relation of ETGs (see Fig.\,\ref{fig:MBH}) with BayesLineFit \citep{Lelli2019}.}
    \label{fig:corner3}
\end{figure}

We present ``corner plots'' from the \texttt{corner.py} package \citep{Corner2016}. The various panels show the posterior probability distribution of pairs
of fitting parameters, and the marginalized probability distribution of each fitting parameter. In the inner panels, individual MCMC samples outside the 2$\sigma$ confidence region are shown with black dots, while binned MCMC samples inside the 2$\sigma$ confidence region are shown by a greyscale; black contours correspond to the 1$\sigma$ and 2$\sigma$ confidence regions; the red squares and solid lines show median values. In the outer panels (histograms), solid and dashed lines correspond to the median and $\pm1\sigma$ values, respectively. Figure\,\ref{fig:corner2} corresponds to the rotation-curve fit for the baseline mass model (Sect.\, \ref{sec:mass}), while Figure\,\ref{fig:corner3} corresponds to linear fits to the $M_\bullet-M_\star$ relation (Sect.\,\ref{sec:coevolution}) using BayesLineFit \citep{Lelli2019}.

% Don't change these lines
\bsp	% typesetting comment
\label{lastpage}
\end{document}